\documentclass[11pt]{iopart}
\pdfoutput=1
\usepackage{fullpage}
\usepackage{iopams,graphicx,color}
\usepackage{setstack}
\usepackage[colorlinks=true]{hyperref} 
\hypersetup{urlcolor=blue,linkcolor=blue,citecolor=violet,colorlinks=true}

\usepackage{color}
\usepackage{pgf,tikz}
\usetikzlibrary{decorations.markings}
\tikzset{>=latex} 
\usepackage{verbatim}
\usepackage{marginnote}
\usepackage{cite}
\usepackage{enumitem}
\usepackage{verbatim,braket,upgreek}
\usepackage{etoolbox}

\usetikzlibrary{fadings,decorations.pathmorphing}

\usepackage{braket}

\begin{document}
 \title{Return probability after a quench from a domain wall initial state in the spin-1/2 XXZ chain}

\author{Jean-Marie St\'ephan$^1$}

\address{
Univ Lyon, Universit\'e Claude Bernard Lyon 1, CNRS UMR 5208, Institut Camille Jordan, 43 blvd. du 11 novembre 1918, F-69622 Villeurbanne cedex, France
}

 \eads{\mailto{stephan@math.univ-lyon1.fr}}
 
\begin{abstract}
We study the return probability and its imaginary ($\uptau$) time continuation after a quench from a domain wall initial state in the XXZ spin chain, focusing mainly on the region with anisotropy $|\Delta|< 1$. We establish exact Fredholm determinant formulas for those, by exploiting a connection to the six vertex model with domain wall boundary conditions. In imaginary time, we find the expected scaling for a partition function of a statistical mechanical model of area proportional to $\uptau^2$, which reflects the fact that the model exhibits the limit shape phenomenon. In real time, we observe that in the region $|\Delta|<1$ the decay for large times $t$ is nowhere continuous as a function of anisotropy: it is either gaussian at root of unity or exponential otherwise. As an aside, we also determine that the front moves as $x_{\rm f}(t)=t\sqrt{1-\Delta^2}$, by analytic continuation of known arctic curves in the six vertex model. 
Exactly at $|\Delta|=1$, we find the return probability decays as $e^{-\zeta(3/2) \sqrt{t/\pi}}t^{1/2}O(1)$. It is argued that this result provides an upper bound on spin transport. In particular, it suggests that transport should be diffusive at the isotropic point for this quench.
\end{abstract}
 \date{\today}
\maketitle

\tableofcontents
\newpage
 \section{Introduction}
Quantum quench problems and statistical mechanical models are related by a mere substitution $\uptau=it$, called Wick rotation, from euclidean time $\uptau$ to real time $t$. This can most easily be seen through the transfer matrix approach to statistical mechanics in $d+1$ dimensions, which selects a spatial direction (euclidean, or imaginary time), along which the transfer matrix is applied. This transfer matrix may be, loosely speaking, interpreted as the exponential of an operator, a Hamiltonian, for a quantum system in $d$ dimension. It turns out the Hamiltonians obtained through this procedure on two-dimensional statistical mechanical models are often relevant physically, in the realm of effectively one-dimensional compounds, or cold atomic gases confined on a one-dimensional line. Examples include the two dimensional Ising model or six vertex models, which map to the quantum Ising chain in transverse field, or anisotropic Heisenberg chain respectively. This point of view is especially useful when studying such systems with field-theoretical methods, where solving the underlying coarse grained statistical model is often simpler than its quantum counterpart.

A special role is played by integrability on both sides of the classical/quantum divide. Integrable models have peculiar features (such as an infinite number of ``local'' conservation laws) that make them exactly solvable. Several physically important observables may be computed exactly, which has been greatly beneficial to the development of both field theory and renormalization \cite{KadanoffCeva,mccoy1973two,di1997conformal}. While this type of models are fine-tuned (they are, strictly speaking, just a set of measure zero in the space of physically relevant models), studying them in detail is still an important subject of research. On the statistical mechanical side the main motivation is universality: several different models, for example at a critical point, belong to the same universality class. Hence integrable models can be good representatives of more realistic physical models. This is not so clear on the non equilibrium side. Integrable system driven violently out of equilibrium, for example after a quantum quench, have peculiar thermalization properties at late times \cite{Cradle,Rigol,Polkovnikov}. Studying them is important, albeit for different reasons. Quantum quench setups are usually realized in cold atomic gases, and many of the corresponding 1d models  turn out be integrable, or close to integrable. These systems are therefore relevant experimentally.

Apart from the obvious fact that the Wick rotation is not justified mathematically in the thermodynamic limit -- it completely ignores Stokes phenomenon -- the above discussion suggests that this procedure sometimes fail for conceptual reasons. This makes is desirable to gain a better understanding of the Wick rotation strategy, in particular determine when it is applicable, and when it is not. There are known examples where this works very well, as is the case of local quenches, where some ground state is perturbed locally. In that case only low energy excitations are generated. The imaginary time approach then allows to make exact predictions regarding light-cone spreading, long time behavior of correlations and entanglement \cite{calabrese2007entanglement,EislerPlatini,stephandubail2011local,calabresecardyreview,DubailStephanVitiCalabrese,DubailStephanCalabrese_curved}. Some of these predictions would be considerably more difficult to study using other approaches. The method has also enjoyed success in more complicated situations such as global quenches, where it for example successfully predicts the linear growth of entanglement \cite{calabrese2007quantum}, despite not capturing the high energy excitations generated after such quenches. 

In this paper, we look at a simple example which may be investigated in depth, while sticking to the logic described above. We start by considering a standard model of statistical mechanics, the six vertex model with domain wall boundary condition, focusing most of our attention on the partition function of the model, which is known exactly \cite{Korepin1982,Izergin1987,IzerginCokerKorepin1992}. A certain limit of this partition function, when Wick-rotated, provides an exact formula for the return probability (RP) after a quantum quench problem in the XXZ spin chain. This particular problem has come under intense scrutiny recently  \cite{antal1999transport,Gobert,MosselCaux,SabettaMisguich,real_time,DubailStephanVitiCalabrese,Prosensuperdiffusion,Prosensuperdiffusion2}. We also explore the relations between the thermodynamic limit of this partition function and the long time behavior of the RP. As we shall see, one does not in general follow from a naive Wick rotation of the other, illustrating the fact that the Wick rotation does not always commute with the thermodynamic limit (surprisingly, it does work at root of unity, however). We will also interpret some of our results in terms of arctic curves (statistical mechanics) and front propagation (quench), providing an example where the Wick rotation procedure does work in the thermodynamic limit. These results will also be discussed in light of recent progress \cite{YoungItalians,Doyonhydro} regarding a hydrodynamic description of such integrable quantum systems out of equilibrium. 

The paper is organized as follows. In section \ref{sec:domainwallquench} we introduce the domain wall quench, and discuss the relation with the six vertex model. We use this to derive an exact formula for the imaginary time RP and the real time RP. The next two sections deal with asymptotics: we look at imaginary times in section \ref{sec:imag}, while real time is the focus of section \ref{sec:real}. We discuss our results and conclude in section \ref{sec:conclusion}.

 \section{Domain wall quench and return probability}
 \label{sec:domainwallquench}
 \subsection{Statement of the problem}
 \label{sec:physics}
We consider the quantum anisotropic Heisenberg spin chain (or XXZ spin chain) on the infinite lattice $\mathbb{Z}+1/2$. The Hamiltonian is given by
\begin{equation}
 H=\sum_{x\in \mathbb{Z}+\frac{1}{2}}\left(S_x^1 S_{x+1}^1+S_x^2 S_{x+1}^2+\Delta\left[S_x^3 S_{x+1}^3-\frac{1}{4}\right]\right),
\end{equation}
with the usual spin operators $S_x^\alpha=\frac{1}{2}\sigma_x^\alpha$, where the $\sigma_x^\alpha$ act as Pauli matrices at site $x$, and as the identity elsewhere. The Hilbert space is $(\mathbb{C}^2)^{\otimes \infty}$, and the spins are, as usual, measured in the basis generated by the eigenstates of the $S_x^3$. The anisotropy is parametrized as 
\begin{equation}
 \cos \gamma=\Delta\quad,\quad 0\leq \gamma\leq \pi.
\end{equation}
Let us consider the following quantum quench protocol. 
The system is initially prepared in the following simple ``domain wall'' product state
\begin{equation}
 \ket{\Psi_0}=\ket{\ldots\uparrow\uparrow\uparrow\uparrow\uparrow\downarrow\downarrow\downarrow\downarrow\downarrow\ldots},
\end{equation}
where all spins at site $x<0$ (resp. $x>0$) are up (resp. down). Then, we let it evolve unitarily with the Hamiltonian $H$, so that the wave function at time $t$ is $\ket{\Psi(t)}=e^{-i H t}\ket{\Psi_0}$. The focus here  is on the ``gapless'' region $|\Delta| \leq 1$, which displays non trivial dynamics at large times (for $\Delta>1$ it freezes after a short while \cite{Gobert,MosselCaux}). 

\begin{figure}[htbp]
\includegraphics[width=8cm]{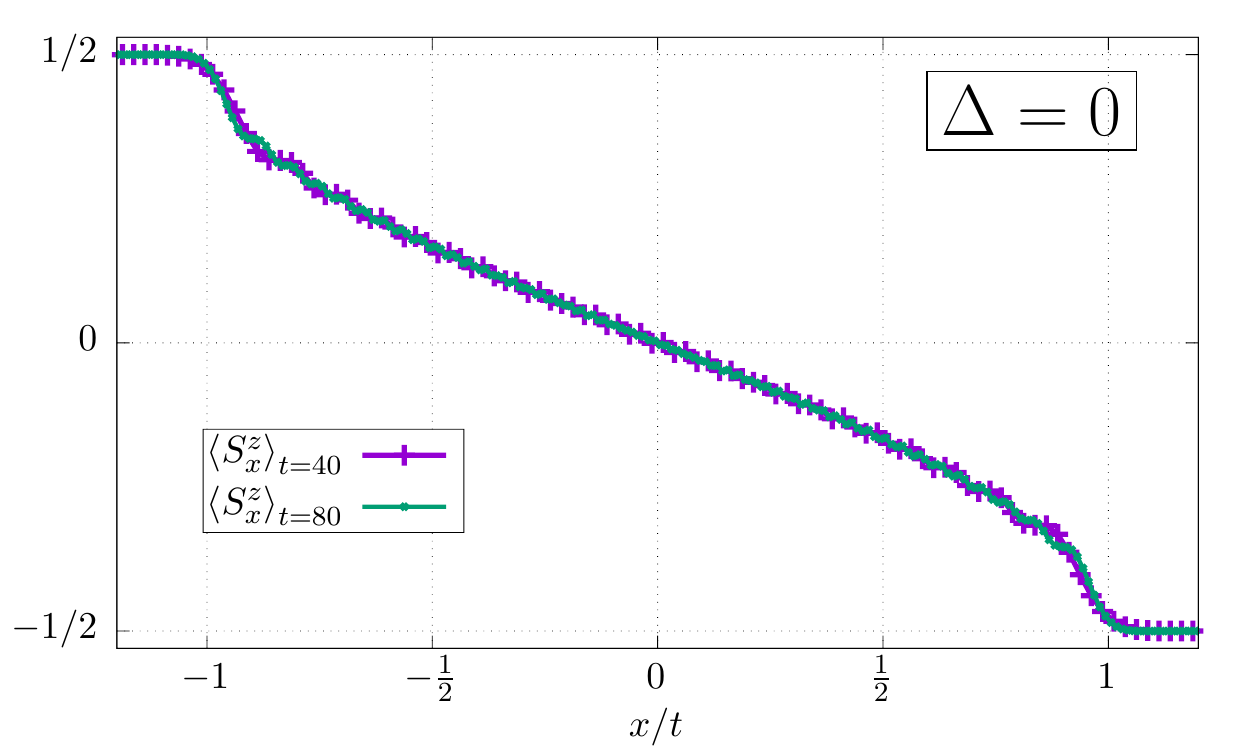}\hfill
\includegraphics[width=8cm]{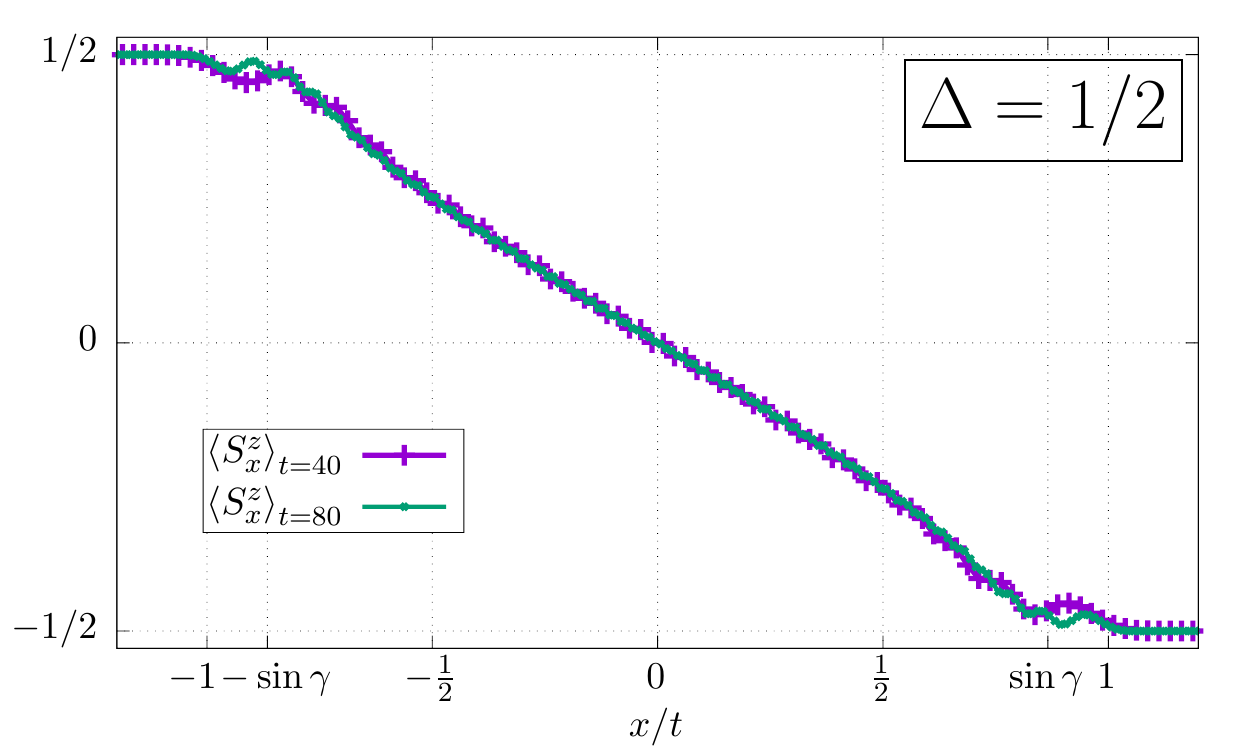}\\
\includegraphics[width=8cm]{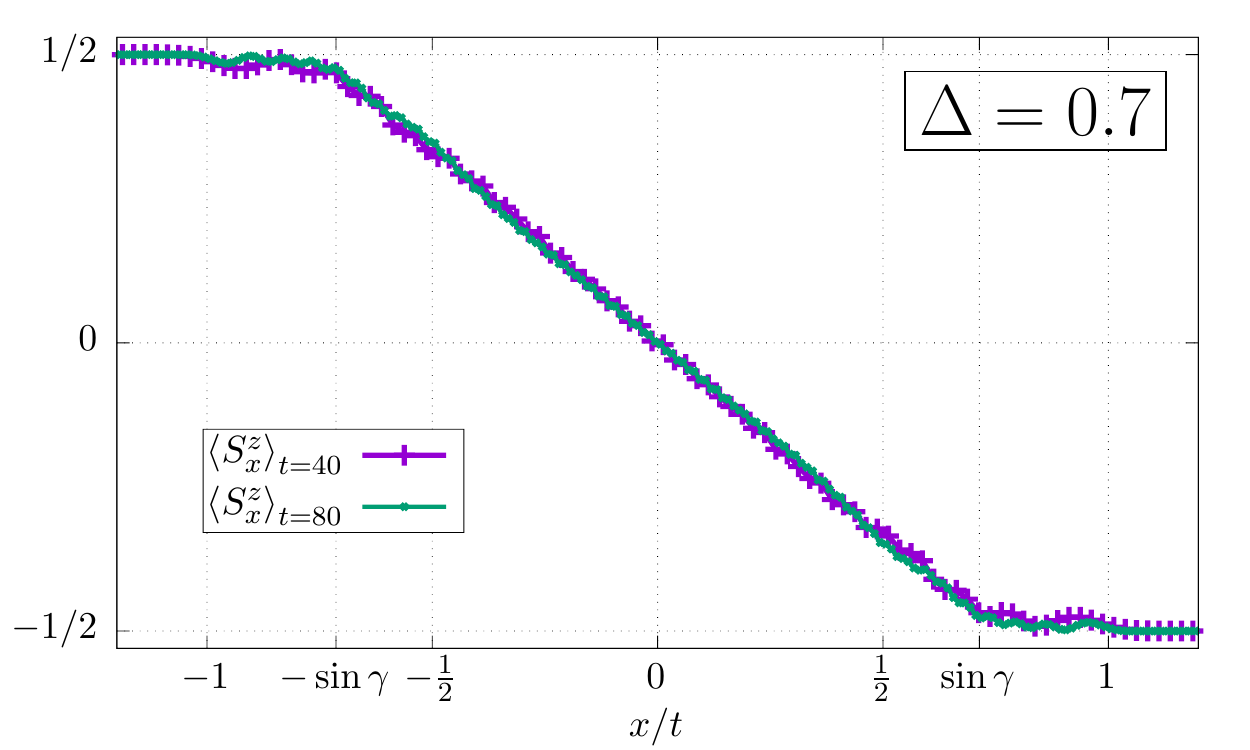}\hfill
\includegraphics[width=8cm]{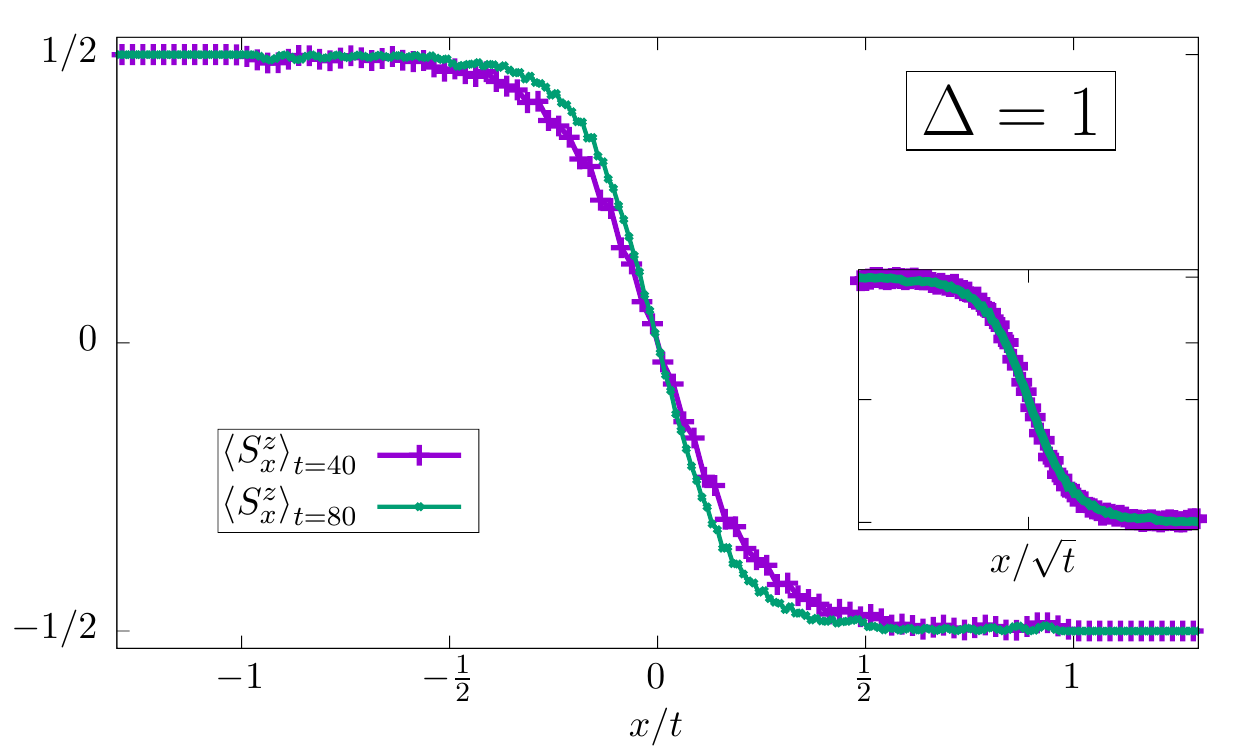}
  \caption{
 tDMRG simulations of the magnetization profile after the quench, shown as a function of $x/t$ for two times $t=40$ and $t=80$. \emph{Top left, top right bottom left:} Anisotropies $\Delta=0,0.5,0.7$, where magnetization appears to converge to a function of $x/t$ when $t\to\infty$. \emph{Bottom right:} Special case $\Delta=1$, where magnetization is not a function of $x/t$ (main plot) but might be a function of $x/\sqrt{t}$ instead (inset).
}
\label{fig:profile}
\end{figure}

Understanding short times properties is fairly easy. Far to the left (right) the initial state looks like an eigenstate of the final Hamiltonian, so no dynamics occurs. The nontrivial behavior takes place near the center $x=0$, where some spin current will flow from left to right. Long time properties have been a subject of intense studies over the last few years \cite{antal1999transport,Gobert,antal2008,Eisler,MosselCaux,SabettaMisguich,real_time,YoungItalians,curvedcft,DubailStephanVitiCalabrese,Prosensuperdiffusion,Prosensuperdiffusion2} (see also Refs.~\cite{EislerDW,VidmarIyerRigol2017} for closely related setups). We summarize here some of those findings, as they will be useful in the following. After the quench correlations have been found to spread ballistically. This can be best illustrated by looking at the magnetization profile at time $t$, for example at the free fermion point $\Delta=0$. In that case magnetization may be expressed in terms of Bessel functions \cite{antal1999transport}, which simplify to
\begin{equation}\label{eq:magn_xx}
 \braket{S_x^3(t)}=-\frac{1}{\pi} \arcsin \frac{x}{t}\qquad,\qquad |x|<t
\end{equation}
in the limit $t\to \infty$ with $x/t$ finite. For $|x|>t$ $\braket{S_x^3(t)}=-\frac{1}{2}{\rm sign} (x/t)$. This defines a front velocity, $v_{\rm f}=1$ here, outside of which magnetization is that of the trivial initial state. This result may also be reproduced by a simpler hydrodynamic argument \cite{antal2008}. Away from free fermions, it has been checked \cite{Gobert,YoungItalians}, that magnetization still remains a function of $x/t$ at large times for all $|\Delta|<1$. We illustrate this in Fig.~\ref{fig:profile}. As can be seen the speed of the front becomes smaller than unity away from $\Delta=0$; in fact we will derive in section~\ref{sec:front} the very simple result $v_{\rm f}= \sin \gamma$. It is important to note that this speed is different from that of the rightmost particle, which lives in a diluted region, and which should be insensitive to interactions, $v_{\rm max}=1$. In the region $\sin \gamma<|x/t|<1$, we observe numerically that $\braket{S_x^3(t)}-\braket{S_x^3(t=0)}$ goes to zero, but the difference in total magnetizations does not. The fact that the two speeds coincide at $\Delta=0$ is a specificity of free fermions. 

Even though the XXZ model is integrable for all values of $\Delta$, exact computations of e.g. magnetization for finite position and time become considerably more involved, so reproducing the path leading to (\ref{eq:magn_xx}) away from free fermions is difficult. A generalized hydrodynamic approach to such problems has been recently pushed forward in Refs.~\cite{Doyonhydro,YoungItalians}. It is different from regular hydrodynamics, as it takes into account the infinite number of conserved charges present in the XXZ chain. Then, solving the corresponding set of Thermodynamic Bethe Ansatz (TBA) equations yields a magnetization profile in excellent agreement with numerical simulations. In the limit $t\to \infty$, the results are expected to become exact. We note that while reasonable and strongly motivated, the hydrodynamic equations of Ref.~\cite{YoungItalians} are a postulate at this point. Proving them starting from the lattice model is a challenging but important problem. While the methods and spirit of the present paper are different (in a sense our aim is less ambitious, as the hydrodynamic approach may be applied to other problems, see e. g. \cite{DoyonDubailKonikYoshimura,Doyon_solitons,Alba_hydroEE,Nonballistic_Italians}), some of our findings will allow us to make contact with this approach. For example we will derive an exact formula for a simpler observable, the return probability. We will also be able to determine exactly the speed of the front, and anticipating on the following, this result is in agreement with the generalized hydrodynamic approach.  
\subsection{Return probability}
\label{sec:returnproba}
 Arguably one of the simplest observable is the return probability (RP)
\begin{equation}
 \mathcal{R}(t)=\left|\mathcal{A}(t)\right|^2\qquad,\qquad \mathcal{A}(t)=\braket{\Psi_0|e^{i H t}|\Psi_0},
\end{equation}
the (modulus square of the) overlap between the wave function at time $t$ and the initial wave function.
 Such a quantity is expected to go to zero very fast, unless the initial state is close to an eigenstate of the final Hamiltonian. For most global quenches the decay takes the form $e^{-L f(t)}$ where $L$ is the system size, see e. g. \cite{Loschmidtreview,LoschmidtZanardi,PozsgayLoschmidt2013,LoschmidtSirker,DynamicalPhaseTransition,Albo}, so that the return probability is zero for any time $t>0$ in an infinite system. This is not so here, as the quench we are studying has mostly local effects.
 
 Let us first mention the case $\Delta>1$, which was studied in Ref.~\cite{MosselCaux} (see also \ref{app:another}). It was shown that  
 $\lim_{t\to \infty} \mathcal{R}(t)=\prod_{p=1}^{\infty}\left(1-e^{-2p \eta}\right)^{2}$ with $\cosh \eta=\Delta$. This result may be explained by the fact that the quench does nothing unless one looks at the center $x=0$ or a few sites away. This is a rare instance of a system not thermalizing after a quench in a many-body system.

 The aim of the present paper is to study the case $|\Delta|\leq 1$, which leads to quite rich behavior. Due to the non zero transport properties after the quench, one expects the RP to go to zero as time increases, but how fast? It turns out there is a simple intuitive guess for the decay, which can be illustrated for $|\Delta|<1$. As already mentioned a front propagates ballistically at some finite speed $v_{\rm f}$. This defines an effective size $v_{\rm f} t$ for the system, outside of which the wave function looks like the initial product state. Since the Hilbert space is made of tensor products of local degrees of freedom, one may then think of the overlap as an overlap between two states with different physical properties (magnetization, current, etc) in a system of effective size $v_{\rm f}t$, which is expected to be exponentially small in system size. Said differently, we expect the (logarithmic) RP to scale linearly with time, $-\log \mathcal{R}(t)=\alpha t+o(t)$, \emph{or faster}. 

The ``or faster'' above is emphasized for a reason, as this intuitive argument might overestimate \interfootnotelinepenalty=10000\footnote{Technically one cannot exclude underestimation either, even though it looks unlikely. Note also that applying this counting argument to overlaps of product states with (say) ground states runs into similar problems in the XXZ chain. For example, the overlap with the N\'eel state can be shown to decay exponentially \cite{BrockmannStephan} with system size $L$, but the overlap with the domain wall state is expected to scale as $e^{-\alpha L^2}$, similar to what happens for the emptiness formation probability \cite{Emptiness1,Emptiness2,Emptiness3,Emptiness4,Emptiness5,Emptiness6}.} the magnitude of the overlap. Indeed, dephasing mechanisms occurring after the quench are in general quite subtle, and extra perfect cancellations cannot be discarded, especially since we are dealing with an integrable system. In fact, this occurs at the only point in the gapless phase where the RP is already known ($\Delta=0$) \cite{prahofer2002scale,weietal,real_time}. In this case the RP is given by the remarkably simple formula $\mathcal{R}(t)=e^{-t^2/4}$, which is gaussian, not exponential. To study this problem away from free fermions, we derive in the next subsection an exact determinant formula. 
 \subsection{Relation to the six vertex model and determinant representation}\label{sec:hamilimit}
 \begin{figure}[htbp]
\centering
 \begin{tikzpicture}
  \draw[thick] (0,0) -- (1,1);
  \draw[thick] (0,1) -- (1,0);
  \draw[line width=3pt,color=blue] (0,0) -- (1,1);
  \draw[line width=3pt,color=blue] (0,1) -- (1,0);
  \draw (0.5,-0.5) node {$a_1$};
  \draw[yshift=-2cm,thick] (0,0) -- (1,1);
   \draw[yshift=-2cm,thick] (0,1) -- (1,0);
  \draw[<-,yshift=-2cm,line width=2.2pt,black] (0.09,0.09) -- (0.18,0.18);
   \draw[<-,yshift=-2cm,line width=2.2pt,black] (0.89,0.11) -- (0.8,0.2);
    \draw[<-,yshift=-1.5cm,xshift=0.5cm,line width=2.2pt,black] (0.12,0.12) -- (0.2,0.2);
   \draw[<-,yshift=-1.5cm,xshift=-0.5cm,line width=2.2pt,black] (0.9,0.1) -- (0.82,0.18);
  \begin{scope}[xshift=2cm]
  \draw[thick] (0,0) -- (1,1);
  \draw[thick] (0,1) -- (1,0);
  \draw (0.5,-0.5) node {$a_2$};
   \draw[yshift=-2cm,thick] (0,0) -- (1,1);
   \draw[yshift=-2cm,thick] (0,1) -- (1,0);
     \draw[->,yshift=-2cm,line width=2.2pt,black] (0.3,0.3) -- (0.39,0.39);
   \draw[->,yshift=-2cm,line width=2.2pt,black] (0.7,0.3) -- (0.61,0.39);
    \draw[->,yshift=-1.5cm,xshift=0.5cm,line width=2.2pt,black] (0.31,0.31) -- (0.4,0.4);
   \draw[->,yshift=-1.5cm,xshift=-0.5cm,line width=2.2pt,black] (0.69,0.31) -- (0.6,0.4);
  \end{scope}
    \begin{scope}[xshift=5cm]
  \draw[thick] (0,0) -- (1,1);
  \draw[thick] (0,1) -- (1,0);
  \draw[line width=3pt,color=blue] (0,0) -- (1,1);
  \draw (0.5,-0.5) node {$b_1$};
   \draw[yshift=-2cm,thick] (0,0) -- (1,1);
   \draw[yshift=-2cm,thick] (0,1) -- (1,0);
    \draw[<-,yshift=-2cm,line width=2.2pt,black] (0.10,0.10) -- (0.19,0.19);
     \draw[<-,yshift=-1.5cm,xshift=0.5cm,line width=2.2pt,black] (0.12,0.12) -- (0.2,0.2);
       \draw[->,yshift=-2cm,line width=2.2pt,black] (0.7,0.3) -- (0.61,0.39);
       \draw[->,yshift=-1.5cm,xshift=-0.5cm,line width=2.2pt,black] (0.69,0.31) -- (0.6,0.4);
  \end{scope}
    \begin{scope}[xshift=7cm]
  \draw[thick] (0,0) -- (1,1);
  \draw[thick] (0,1) -- (1,0);
  \draw[line width=3pt,color=blue] (0,1) -- (1,0);
  \draw (0.5,-0.5) node {$b_2$};
   \draw[yshift=-2cm,thick] (0,0) -- (1,1);
   \draw[yshift=-2cm,thick] (0,1) -- (1,0);
   \draw[<-,yshift=-2cm,line width=2.2pt,black] (0.89,0.11) -- (0.8,0.2);
   \draw[<-,yshift=-1.5cm,xshift=-0.5cm,line width=2.2pt,black] (0.88,0.12) -- (0.8,0.2);
     \draw[->,yshift=-2cm,line width=2.2pt,black] (0.3,0.3) -- (0.39,0.39);
    \draw[->,yshift=-1.5cm,xshift=0.5cm,line width=2.2pt,black] (0.31,0.31) -- (0.4,0.4);
  \end{scope}
    \begin{scope}[xshift=10cm]
  \draw[thick] (0,0) -- (1,1);
  \draw[thick] (0,1) -- (1,0);
  \draw[line width=3pt,color=blue] (0,0) -- (0.5,0.5);
  \draw[line width=3pt,color=blue] (0,1) -- (0.5,0.5);
  \draw (0.5,-0.5) node {$c_1$};
   \draw[yshift=-2cm,thick] (0,0) -- (1,1);
   \draw[yshift=-2cm,thick] (0,1) -- (1,0);
   \draw[<-,yshift=-2cm,line width=2.2pt,black] (0.10,0.10) -- (0.19,0.19);
   \draw[<-,yshift=-1.5cm,xshift=-0.5cm,line width=2.2pt,black] (0.88,0.12) -- (0.8,0.2);
   \draw[->,yshift=-2cm,line width=2.2pt,black] (0.7,0.3) -- (0.61,0.39);
    \draw[->,yshift=-1.5cm,xshift=0.5cm,line width=2.2pt,black] (0.31,0.31) -- (0.4,0.4);
  \end{scope}
    \begin{scope}[xshift=12cm]
  \draw[thick] (0,0) -- (1,1);
  \draw[thick] (0,1) -- (1,0);
  \draw[line width=3pt,color=blue] (1,0) -- (0.5,0.5);
  \draw[line width=3pt,color=blue] (1,1) -- (0.5,0.5);
  \draw (0.5,-0.5) node {$c_2$};
   \draw[yshift=-2cm,thick] (0,0) -- (1,1);
   \draw[yshift=-2cm,thick] (0,1) -- (1,0);
        \draw[->,yshift=-2cm,line width=2.2pt,black] (0.3,0.3) -- (0.39,0.39);
   \draw[->,yshift=-1.5cm,xshift=-0.5cm,line width=2.2pt,black] (0.69,0.31) -- (0.6,0.4);
   \draw[<-,yshift=-2cm,line width=2.2pt,black] (0.89,0.11) -- (0.8,0.2);
    \draw[<-,yshift=-1.5cm,xshift=0.5cm,line width=2.2pt,black] (0.12,0.12) -- (0.2,0.2);
  \end{scope}
 \end{tikzpicture}
\caption{Weights of the six vertex model. The vertex configurations are shown at the bottom. \emph{Top:} equivalent representation in terms of directed self-avoiding trajectories (thick blue lines), which we use in the following. We also choose the weights to be invariant under a global reversal of all arrows. This implies $a_1=a_2=a$, $b_1=b_2=b$, $c_1=c_2=c$.}
\label{fig:6vweights}
\end{figure}
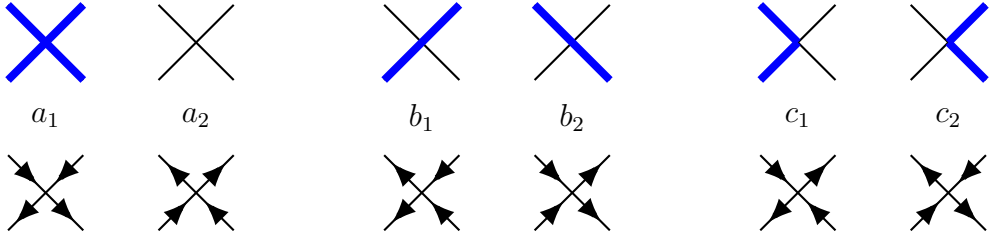
An exact formula for the RP can be obtained using known results on a closely related model, the two-dimensional classical six-vertex model \cite{baxter}. Let us specify the conventions we will use. The weights of the six vertex model are denoted by $a_1,a_2$, $b_1,b_2$ and $c_1,c_2$ (see figure \ref{fig:6vweights}). We look at the special case, $a_1=a_2=a$, $b_1=b_2=b$, $c_1=c_2=c$. The corresponding anisotropy in $XXZ$ is $\Delta=\frac{a^2+b^2-c^2}{2ab}$.

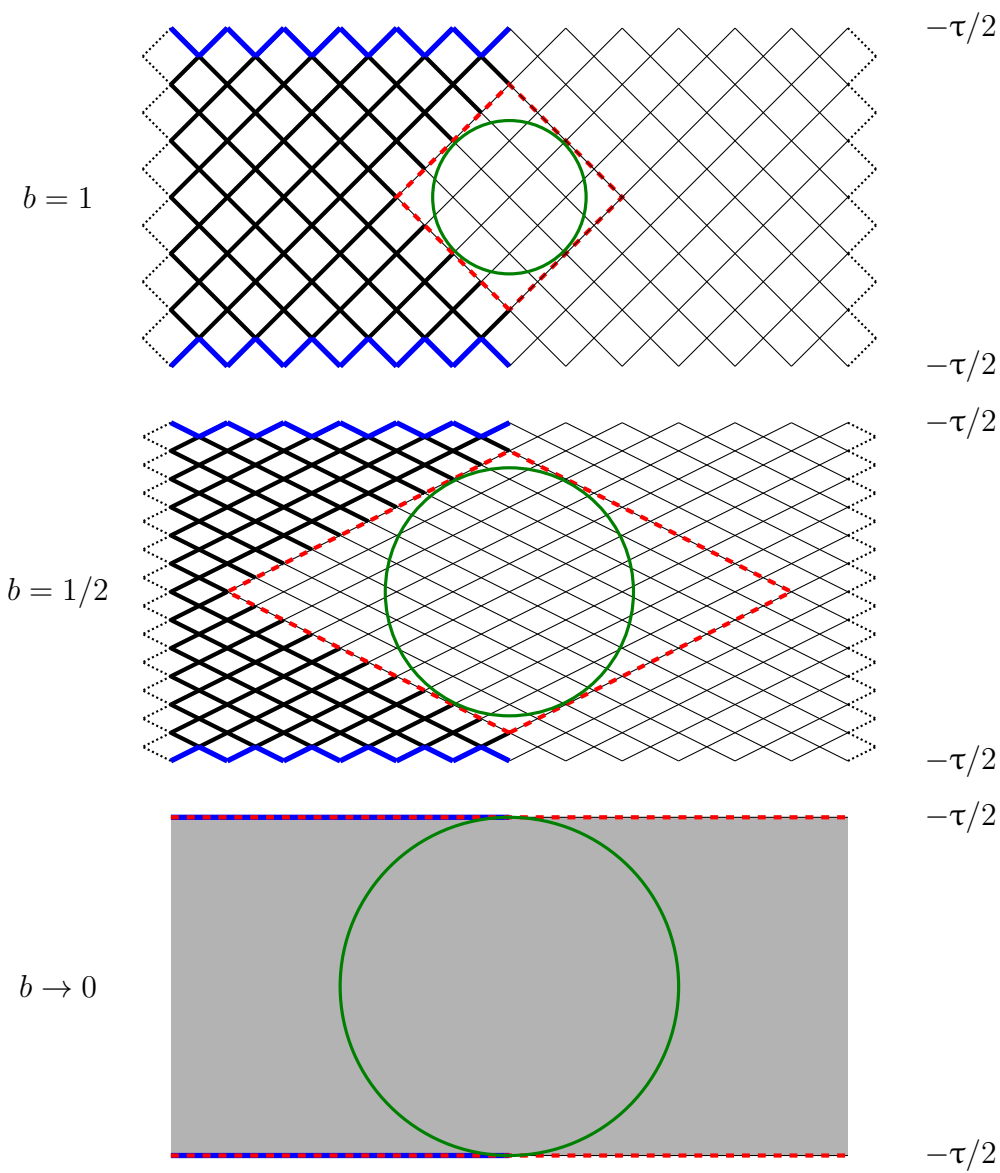
\begin{figure}[htbp]
 \centering\begin{tikzpicture}[scale=1.5]
 \draw (7,0) node {$-\uptau/2$};
  \draw (7,3) node {$-\uptau/2$};
 \draw (-1,1.5) node {$b=1$};
  \draw[thin] (0,0) -- (3,3);
  \foreach \x in {1,2,3,4,5}{
  \draw[thin] (0.5*\x,0) -- (3,3-0.5*\x);
  \draw[thin] (0,0.5*\x) -- (3-0.5*\x,3);
  }
  \draw[thin] (0,3) -- (3,0);
  \foreach \x in {1,2,3,4,5}{
  \draw[thin] (3,0.5*\x) -- (0.5*\x,3);
  \draw[thin] (0.5*\x,0) -- (0,0.5*\x);
  }
\foreach \y in {0,1,2,3,4,5}{
\draw[thick,densely dotted] (0,0.5*\y) -- (-0.25,0.5*\y+0.25);
\draw[thick,densely dotted] (0,0.5*\y+0.5) -- (-0.25,0.5*\y+0.25);
\draw[thick,densely dotted] (6,0.5*\y) -- (6.25,0.5*\y+0.25);
\draw[thick,densely dotted] (6,0.5*\y+0.5) -- (6.25,0.5*\y+0.25);
}
\foreach \x in {0,1,2,3,4,5}{
\draw[black,ultra thick] (2.75-0.5*\x,0.25) -- (0,3-0.5*\x);
\draw[black,ultra thick] (0,0.5*\x) -- (3-0.5*\x,3);
}
\foreach  \x in {0,1,2,3,4}{
\draw[black,ultra thick] (0.75+0.5*\x,0.25) -- (2+0.25*\x,1.5-0.25*\x);
\draw[black,ultra thick] (0.75+0.5*\x,2.75) -- (2+0.25*\x,1.5+0.25*\x);
}
\foreach \x in {0,1,2,3,4,5}{
\draw[blue,line width=2pt] (0.5*\x,0) -- (0.25+0.5*\x,0.25);
\draw[blue,line width=2pt] (0.25+0.5*\x,0.25) -- (0.5+0.5*\x,0);
\draw[blue,line width=2pt] (0.5*\x,3) -- (0.25+0.5*\x,2.75);
\draw[blue,line width=2pt] (0.25+0.5*\x,2.75) -- (0.5+0.5*\x,3);
}
\draw[ultra thick,red,dashed] (3,0.5) -- (2,1.5) -- (3,2.5) -- (4,1.5) -- cycle;
\begin{scope}[xshift=3cm]
   \draw (0,0) -- (3,3);
  \foreach \x in {1,2,3,4,5}{
  \draw (0.5*\x,0) -- (3,3-0.5*\x);
  \draw (0,0.5*\x) -- (3-0.5*\x,3);
  }
  \draw (0,3) -- (3,0);
  \foreach \x in {1,2,3,4,5}{
  \draw (3,0.5*\x) -- (0.5*\x,3);
  \draw (0.5*\x,0) -- (0,0.5*\x);
  }
\foreach \x in {0,1,2}{
}
\end{scope}
  \draw[very thick,green!50!black] (3,1.5) circle (0.68);

\begin{scope}[yshift=-3.5cm]
 \draw (7,0) node {$-\uptau/2$};
  \draw (7,3) node {$-\uptau/2$};
 \draw (-1,1.5) node {$b=1/2$};
 \draw (0,0) -- (6,3);
 \foreach \x in {1,2,3,4,5,6,7,8,9,10,11}{
  \draw (0.5*\x,0) -- (6,3-0.25*\x);
  \draw (0,0.25*\x) -- (6-0.5*\x,3);
  }
  \draw (0,3) -- (6,0);
  \foreach \x in {1,2,3,4,5,6,7,8,9,10,11}{
  \draw (6,0.25*\x) -- (0.5*\x,3);
  \draw (0.5*\x,0) -- (0,0.25*\x);
  }

  \foreach \x in {0,1,2,3,4,5}{
\draw[black,ultra thick] (2.75-0.5*\x,0.125) -- (0,1.5-0.25*\x);
\draw[black,ultra thick] (0,1.5+0.25*\x) -- (3-0.5*\x,3);
}
\foreach  \x in {0,1,2,3,4,5}{
\draw[black,ultra thick] (0.25+0.5*\x,0.125) -- (1.75+0.25*\x,0.875-0.125*\x);
\draw[black,ultra thick] (0.25+0.5*\x,2.875) -- (1.75+0.25*\x,2.125+0.125*\x);
}
\foreach \x in {1,2,3,4,5}{
\draw[black,ultra thick] (0,0.25*\x) -- (1.75-0.25*\x,0.875+0.125*\x);
\draw[black,ultra thick] (0,3-0.25*\x) -- (1.75-0.25*\x,2.125-0.125*\x);
}
\foreach \y in {0,1,2,3,4,5,6,7,8,9,10,11}{
\draw[thick,densely dotted] (0,0.25*\y) -- (-0.25,0.25*\y+0.125);
\draw[thick,densely dotted] (0,0.25*\y+0.25) -- (-0.25,0.25*\y+0.125);
\draw[thick,densely dotted] (6,0.25*\y) -- (6.25,0.25*\y+0.125);
\draw[thick,densely dotted] (6,0.25*\y+0.25) -- (6.25,0.25*\y+0.125);
}

  \foreach \x in {0,1,2,3,4,5}{
\draw[blue,line width=2pt] (0.5*\x,0) -- (0.25+0.5*\x,0.125);
\draw[blue,line width=2pt] (0.25+0.5*\x,0.125) -- (0.5+0.5*\x,0);
\draw[blue,line width=2pt] (0.5*\x,3) -- (0.25+0.5*\x,2.875);
\draw[blue,line width=2pt] (0.25+0.5*\x,2.875) -- (0.5+0.5*\x,3);
}
\draw[ultra thick,red,dashed] (3,0.25) -- (0.5,1.5) -- (3,2.75) -- (5.5,1.5) -- cycle;
\draw[very thick,green!50!black] (3,1.5) circle (1.1);
\end{scope}

\begin{scope}[yshift=-7cm]
 \draw (7,0) node {$-\uptau/2$};
  \draw (7,3) node {$-\uptau/2$};
 \draw (-1,1.5) node {$b\to 0$};
\fill[color=black!30] (0,0) -- (6,0) -- (6,3) -- (0,3) -- cycle;
\draw (0,0) -- (6,0);
\draw (0,3) -- (6,3);
 \draw[blue,line width=2pt] (0,0) -- (3,0);
 \draw[blue,line width=2pt] (0,3) -- (3,3);
 \draw[ultra thick,red,dashed] (0,0) -- (6,0);
 \draw[ultra thick,red,dashed] (0,3) -- (6,3);
 \draw[very thick,green!50!black] (3,1.5) circle (1.5);
 \end{scope}
 \end{tikzpicture}
 \caption{Illustration of the Hamiltonian limit of the six vertex model with domain wall boundary conditions. \emph{Top:} Six vertex with the domain wall state (shown in thick blue) imposed at the top and bottom (the horizontal direction is infinite). Due to the ice rules, some vertices are automatically set as a consequence (shown in thick black). Therefore, only the vertices inside the red dashed lozenge can fluctuate. This generates the six-vertex model with domain wall boundary conditions. From top to bottom $b=1$, $b=1/2$, $b\to 0$. The number of vertical steps is taken to be proportional to $1/b$, but the distance $\uptau$ between the top and bottom lines is kept constant. We also set $a=1$. In the thermodynamic limit ($\uptau\to\infty$) vertices may fluctuate only inside a certain region, the boundary of which is called an arctic curve. \emph{Green circles:} What happens to the ``arctic circle'' $x^2+y^2=(\uptau/2)^2/(1+b^2)$, at $\Delta=0$, in the thermodynamic limit.}
 \label{fig:hamlimit}
\end{figure}

The quantity $\mathcal{Z}(\uptau)=\braket{\Psi_0|e^{H \uptau}|\Psi_0}$ for real $\uptau>0$ is related to the  partition function of the six vertex model with domain wall boundary conditions (DWBC; note the unfortunate clash of terminology here, as DWBC refers to the boundary conditions shown in dashed red in figure~\ref{fig:hamlimit}, they are different from the domain wall product state, which corresponds to the top and bottom vertices in blue). The idea is to express the partition function using the transfer matrix formalism, as was done in Ref.~\cite{curvedcft} (see figure \ref{fig:hamlimit}).
We have
\begin{equation}
 \braket{\Psi_0|\mathcal{T}^{\,2n}|\Psi_0}=Z_n(a,b,\Delta),
\end{equation}
where $Z_n(a,b,\Delta)$ is the partition function of the six vertex model with domain wall boundary conditions, and $\mathcal{T}$ denotes the transfer matrix of the six vertex model. 
This partition function was first considered by Korepin \cite{Korepin1982}, in relation with Gaudin's formula for the norm of the Bethe states. Later, Izergin \cite{Izergin1987,IzerginCokerKorepin1992} derived an elegant  determinant formula, which reads in the homogeneous limit
\begin{equation}\label{eq:izergin}
 Z_n(a,b,\Delta)=\frac{\left[d \sin \epsilon \sin (\gamma+\epsilon)\right]^{n^2}}{\prod_{k=0}^{n-1}k!^2}\det_{0\leq i,j \leq n-1}
 \left(\int_{-\infty}^{\infty}du \,u^{i+j}e^{-\epsilon u}\frac{1-e^{-\gamma u}}{1-e^{-\pi u}}\right),
\end{equation}
with parametrization of the weights as follows
\begin{equation}\fl
 a= d \sin (\gamma+\epsilon)\quad,\quad b=d \sin \epsilon\quad,\quad c=d\sin \gamma\quad,\quad \Delta=\frac{a^2+b^2-c^2}{2ab} =\cos \gamma.
\end{equation}
As is well known, the Hamiltonian $H$ of the XXZ spin chain is essentially the first order term in the expansion of the transfer matrix around $b=0$. Since $e^{\uptau H}=\lim_{n\to \infty}(1+\frac{\uptau H}{n})^n$, one can show that $\mathcal{Z}(\uptau)$ follows from the Trotter (or Hamiltonian) limit
\begin{equation}\label{eq:hamlimit}
 \mathcal{Z}(\uptau)=\braket{e^{\uptau H}}=\lim_{n\to \infty} Z_n(a=1,b=\frac{\uptau}{2n},\Delta).
\end{equation}
We note a similar calculation has also been performed for the N\'eel state \cite{PozsgayLoschmidt2013,LoschmidtXXZNeel}. 
Some care must be taken when taking the limit (\ref{eq:hamlimit}) in Eq.~(\ref{eq:izergin}), as the size of the determinant goes to infinity in the process. This problem may be circumvented by transforming the finite determinant of size $n$ in the Fredholm determinant of an operator, in which $n$ appears only as a parameter. It was, in fact, already demonstrated how this can be achieved in Ref.~\cite{Slavnov} (see also \cite{ColomoPronkodet}, which leads to an alternative formula given in \ref{app:another}), using orthogonal polynomial techniques \cite{Szegobook}. After that the desired limit can be safely taken. We find the Fredholm determinant formula
\begin{equation}\label{eq:fred}
 \mathcal{Z}(\uptau)=\braket{e^{\uptau H}}=e^{-\frac{1}{24}(\uptau \sin \gamma)^2}\det(I-V),
\end{equation}
where $V$ is an operator with kernel
\begin{equation}
 V(x,y)=B_0(x,y)\, \omega(y).
\end{equation}
Here $B_0$ is a special case of the Bessel kernel $B_\alpha$, known in the context of random matrix theory. It is given by
\begin{equation}
 B_\alpha(x,y)=\frac{\sqrt{y}J_\alpha(\sqrt{x})J_\alpha'(\sqrt{y})-\sqrt{x}J_\alpha(\sqrt{y})J_\alpha'(\sqrt{x})}{2(x-y)},
\end{equation}
where $J_\alpha$ is a Bessel function of the first kind, and $J_\alpha'$ its derivative. The function $\omega(y)$ is given by
\begin{equation}
 \omega(y)=\Theta(y)-\frac{1-e^{-\gamma y/(2\uptau\sin \gamma)}}{1-e^{-\pi y/(2\uptau \sin \gamma)}}.
\end{equation}
The kernel $V$ acts on $L^2(\mathbb{R})$.
We refer to \ref{app:fredholmderivation} for a derivation of this result. To get the return probability, it then suffices to set $\uptau=it$. We also recall the definition of the Fredholm determinant of an operator $I-V$:
\begin{equation}
 \det(I-V)=\sum_{k=0}^\infty
 \frac{(-1)^k}{k!} \int_\mathbb{R} dx_1\ldots \int_{\mathbb{R}}dx_k\, \det_{1\leq i,j\leq k} V(x_i,x_j).
\end{equation}
In the remainder of the paper, we explore the asymptotic behavior of our exact result (\ref{eq:fred}) for large imaginary (section~\ref{sec:imag}) and real times (section~\ref{sec:real}). Before doing that let us make two remarks. First, the kernel has a peculiar structure, of the form $[F(x)G(y)-G(x)F(y)]/(x-y)$. The corresponding integral operators are integrable, in the sense of Ref.~\cite{IIKS}. It is possible to associate a Riemann-Hilbert problem to them \cite{IIKS}, which is usually a good starting point for a rigorous asymptotic analysis \cite{DeiftZhou}. Second, Fredholm determinants can be evaluated numerically. We discuss in \ref{app:fredholmnumerical} two different ways of achieving this.
 \section{Asymptotic results in imaginary time and arctic curves}
 \label{sec:imag}
 \subsection{Area scaling and arctic curves}
 \label{sec:arcticcurves}
 We recall here the known result \cite{KorepinZinn,Zinn,BleherFokin} for the asymptotics of the aforementioned six-vertex partition function (we set $a=1$, which is the only case relevant to us):
\begin{equation}\label{eq:6vdwbc}
 Z_n \,\underset{n\to\infty}{\sim}\, e^{n^2\log\left(\frac{\pi \sin \epsilon}{\delta\sin (\pi \epsilon/\delta)}\right)}(\epsilon n)^{\kappa(\gamma)}\,O(1),
\end{equation}
where $\delta=\pi-\gamma$. The power law exponent is given by
 \begin{equation}
   \kappa(\gamma)=\frac{1}{12}-\frac{(\pi-\gamma)^2}{6\pi \gamma} . 
 \end{equation}
The leading gaussian term in (\ref{eq:6vdwbc}) was obtained by Korepin and Zinn-Justin \cite{KorepinZinn,Zinn} using Toda chain hierarchies
and random matrix techniques. The power-law exponent was derived rigorously by Bleher and Fokin \cite{BleherFokin}, using the Riemann-Hilbert approach to Hankel determinants. 

 The results mentioned above may be used to obtain the large $\uptau$ behavior of $\mathcal{Z}(\uptau)$, simply by performing the Hamiltonian limit in the expansion (\ref{eq:6vdwbc}): 
 \begin{equation}\label{eq:imag_as}
 \mathcal{Z}(\uptau)\,\underset{\uptau \to \infty}{\sim}\, \exp\left(\left[\frac{\pi^2}{(\pi-\gamma)^2}-1\right]\frac{(\uptau \sin \gamma)^2}{24}\right)\uptau^{\kappa(\gamma)} O(1).                                                                                                                                                                                 \end{equation}
We have of course assumed that the Hamiltonian limit commutes with the large $n$ limit in the six vertex model, which is not guaranteed a priori. However, the arguments of Refs.~\cite{KorepinZinn,Zinn} for the leading term can easily be adapted to situation where $b\sim n^{-\nu}$ for some $\nu<1$, leading to the same result. The fact that the same power law exponent occurs in (\ref{eq:6vdwbc}) and (\ref{eq:imag_as}) may be justified by invoking a universality argument.

\begin{figure}[htbp]
 \centering\includegraphics[width=12cm]{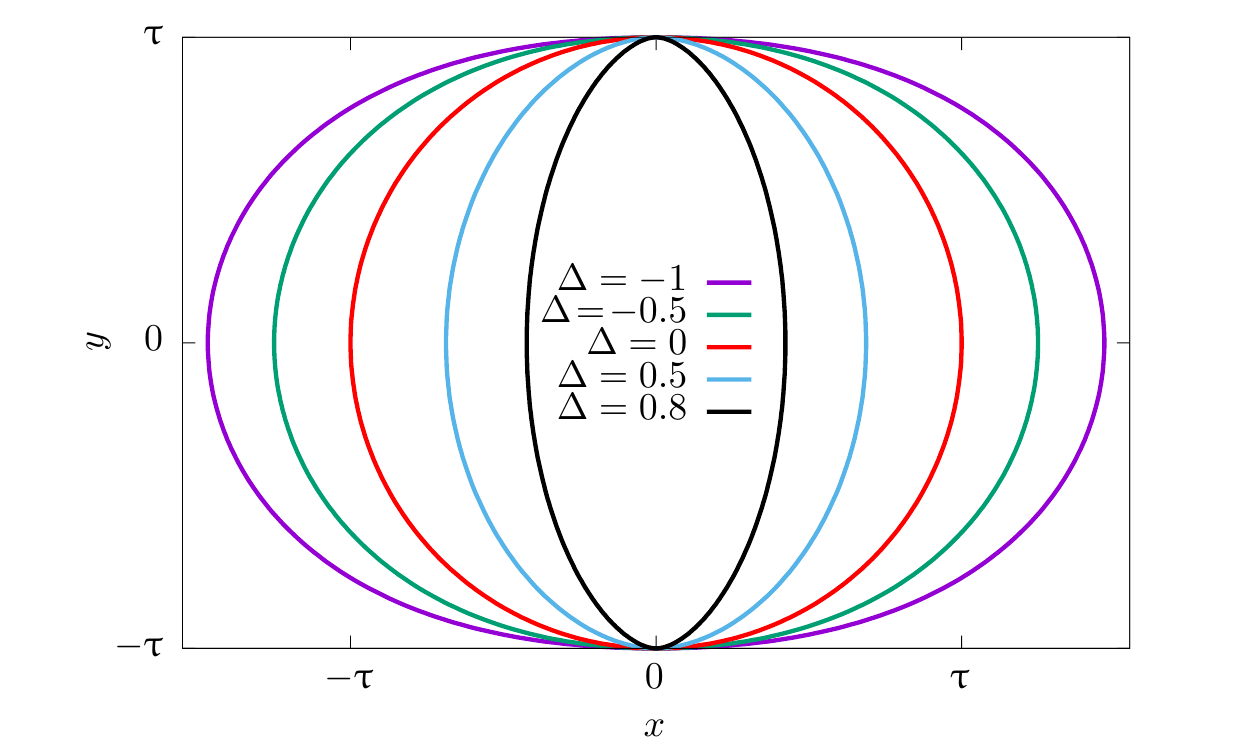}
 \caption{Arctic curves for several values of $\Delta$ in the XXZ chain. The interior is fluctuating, the exterior freezes in the thermodynamic limit. The horizontal width shrinks to zero as $\sqrt{1-\Delta}$ in the limit $\Delta\to 1$.}
 \label{fig:arctic_curves}
\end{figure}

The result has a simple physical interpretation, which follows from the limit shape phenomenon. There is a spatial phase separation in the six-vertex model with domain wall boundary conditions: the vertices of the model only fluctuate inside a certain well defined region in the thermodynamic limit. Outside of this region the vertices are completely frozen; the curve separating the two phases is called the arctic curve. For $\Delta=0$ it has been proven to be a circle \cite{jockusch1998random}. Colomo and Pronko \cite{ColomoPronkocurve} conjectured an explicit formula for general $|\Delta|<1$ (see also \cite{ColomoPronkoZinn} for other values of $\Delta$). We checked that the Hamiltonian limit of their result is well-defined. We find, after long algebra, the following result
\small
\begin{eqnarray}\fl\nonumber
 \frac{x}{\uptau}&=&\frac{\sin s  \sin (\gamma +s ) \left[\alpha ^2 \csc ^2\alpha s  \left\{\cos (2 \gamma +3 s ) (\cos s -\alpha  \sin s  \cot \alpha  s)+\alpha  \sin s  \cos s \cot \alpha  s+\cos ^2 s -2\right\}+2\right]}{\sin ^2(\gamma +s )+\sin ^2 s },\\	\fl \label{eq:arcticcurve}
\frac{y}{\uptau} &=&\frac{\sin ^2(\gamma +s ) \left[2 \alpha ^2 \csc \gamma  \sin ^2 s  \csc ^2\alpha  s  \left\{ 2 \alpha  \sin s  \cot \alpha  s  \sin (\gamma +s )-\sin (\gamma +2 s)\right\}-1\right]+\sin ^2s }{\sin ^2(\gamma +s )+\sin ^2 s },
\end{eqnarray}
\normalsize
with $s\in [0;\gamma]$ and $\alpha=\pi/(\pi-\gamma)$. This parametric equation only describes the left part of the curve, the right part follows from the $x\to -x$ symmetry. Note that, similar to the six vertex case \cite{ColomoNoferiniPronko}, it becomes algebraic at root of unity. For example, we recover the circle $x^2+y^2=(\uptau/2)^2$ of Refs~\cite{prahofer2002scale,curvedcft} at $\Delta=0$, and an algebraic curve of degree $5$ (resp. $9$) at $\Delta=-1/2$ (resp. $\Delta=1/2$). The exact expression is perhaps not particularly illuminating, but it will be useful in section~\ref{sec:front}. The curve is shown in Fig.~\ref{fig:arctic_curves} for several values of $\Delta$. 

Now, $\log \mathcal{Z}(\uptau)$ may be interpreted as the free energy of the fluctuating region, which should  scale as the area $\uptau^2$ of this region, hence justifying (\ref{eq:imag_as}). The width of the domain shrinks to zero when $\Delta\to 1$ ($\gamma\to 0$), consistent with the fact the coefficient of $\uptau^2$ in the exponential (\ref{eq:imag_as}) goes to zero in that limit. The scaling of the partition function for $\Delta=1$ is studied in the next subsection. 
 \subsection{The point $\Delta=1$}
 \label{sec:isotropic}
 The point $\Delta=1$ ($\gamma=0$) is special, and requires a different analysis. The exact formula (\ref{eq:fred}) takes a simpler form in that case, 
 \begin{equation}\label{eq:fred_delta1}
  \mathcal{Z}(\uptau)=\det(I-V)\qquad,\qquad 
 V(x,y)=B_0(x,y)e^{-\frac{x+y}{4\uptau}},
 \end{equation}
where the kernel $V$ acts on $L^2(\mathbb{R}_+)$. We note also a close relation \footnote{I am grateful to Pavel Krapivsky and Kirone Mallick for pointing out that similarity to me.} with the symmetric simple exclusion process (SSEP) \cite{Spohnbook} and related growth problems. In fact, (\ref{eq:fred_delta1}) coincides with an exact large deviation result of Ref.~\cite{DerridaGerschenfeld2009} (in the limit of vanishing current, see \ref{app:alternative}).

It is possible to evaluate the large $\uptau$ behavior of each ${\rm Tr}\, V^k$, and recover the large $\uptau$ behavior of $\mathcal{Z}(\uptau)$ through the identity $\log \det(I-V)=-\sum_{k=1}^{\infty}\frac{{\rm Tr}\, V^k}{k}$. The leading term is known from \cite{DerridaGerschenfeld2009,KrapivskyMallickSadhu2015}, and may be reproduced from the following heuristic argument. First, a change of variables shows $\det(I-V)=\det(I-\tilde{V})$, with kernel $\tilde{V}(x,y)=2\uptau \sqrt{xy} V(\uptau x^2,\uptau y^2)$. Then, using the asymptotic expansion of the Bessel function for large $\uptau$, one is lead to consider a simpler version of our partition function, $\det(I-W)$, which should have the same leading asymptotic behavior as $\det(I-V)$. The point is the kernel $W$ takes a  simple form
\begin{equation}
 W(x,y)=\frac{\sin \sqrt{\uptau}(x-y)}{\pi(x-y)}e^{-\frac{x^2+y^2}{4}},
\end{equation}
namely a sine kernel confined in a harmonic potential. The trace of $W^k$ reads
\begin{equation}\fl
 {\rm Tr} \, W^k=\int dx_1 \ldots dx_k \frac{\sin \sqrt{\uptau}(x_1-x_2)}{\pi(x_1-x_2)} \frac{\sin \sqrt{\uptau}(x_2-x_3)}{\pi(x_2-x_3)} \ldots \frac{\sin \sqrt{\uptau}(x_k-x_1)}{\pi(x_k-x_1)} e^{-\frac{1}{2}(x_1^2+\ldots +x_k^2)},
\end{equation}
where each integral runs over $\mathbb{R}_+$. 
For large $\uptau$ this is a highly oscillatory integral, which should be dominated by the region where all integration variables are close to each other. With this in mind it is natural to make the change of variables $y_1=x_1-x_2$, $y_2=x_2-x_3$, \ldots, $y_{k-1}=x_{k-1}-x_k$, and ``center of mass'' $y_k=(x_1+\ldots +x_k)/k$. The factors are chosen so that the Jacobian is one. Extending the integration domain of the variables $x_1$,\ldots $x_{k-1}$ to $\mathbb{R}$, and performing the gaussian integral over the center of mass variable, we obtain
 \begin{equation}\fl
  {\rm Tr}\, W^k \sim \sqrt{\frac{\pi}{2k}}\int_{\mathbb{R}^{k-1}}dy_1\ldots dy_{k-1}
  \frac{\sin \sqrt{\uptau} y_1}{\pi y_1}\ldots  \frac{\sin \sqrt{\uptau} y_{k-1}}{\pi y_{k-1}}
   \frac{\sin \sqrt{\uptau} (y_1+\ldots y_{k-1})}{\pi (y_1+\ldots y_{k-1})}e^{P(y_1,\ldots,y_{k-1})}.
 \end{equation}
For large $\uptau$ the rhs may be evaluated by changing variables to $\tilde{y}_k=y_k/\sqrt{\uptau}$, using the fact that $P(0,\ldots,0)=0$, and repeated use of the identity $\int_\mathbb{R} dx \frac{\sin u}{\pi u}\frac{\sin(u+v)}{\pi(u+v)}=\frac{\sin v}{\pi v}$. Hence
\begin{equation}\label{eq:tracek}
 {\rm Tr}\, V^k \sim \sqrt{\frac{\uptau}{2\pi k}}
\end{equation}
for large $\uptau$, which gives \cite{DerridaGerschenfeld2009} $\det(I-V)\approx e^{-\zeta(3/2)\sqrt{\uptau/2\pi}}$, where $\zeta(s)=\sum_{n=1}^\infty n^{-s}$ is the Riemann zeta function. It is possible to refine this asymptotic result, by computing the additive order one contribution to  (\ref{eq:tracek}). This can be done by a more careful treatment of the multidimensional integrals, as well as the error terms coming from the approximation of the kernel $V$ by $W$. We find, after a long calculation,
\begin{equation}
  {\rm Tr}\, V^k =\sqrt{\frac{\uptau}{2\pi k}}-\frac{1}{4\pi}\sum_{p=1}^{k-1}\frac{1}{\sqrt{p(k-p)}}+o(1).
\end{equation}
The second contribution goes to $-\frac{1}{4}$ as $k$ goes to infinity, due to the identity $\int_{0}^k \frac{du}{\sqrt{u(k-u)}}=\pi$. Naively this gives a divergent contribution to $\log \det(I-V)$; however, a natural cutoff to the series is provided by $\uptau$. This leads to the main result of the subsection
\begin{equation}\label{eq:iso_asym}
 \log \mathcal{Z}(\uptau)=-\zeta(3/2)\sqrt{\frac{\uptau}{2\pi}}+\frac{1}{4}\log \uptau +O(1).
\end{equation}
Presumably, subleading corrections take the form of a power series in $\uptau^{-1/2}$. The determinant may also be evaluated numerically for large $\uptau$, using the method presented in \ref{sec:quadrature}. We used this to check formula (\ref{eq:iso_asym}) with very high precision. For example, a fit of $\log[ \mathcal{Z}(\uptau)e^{\frac{\zeta(3/2)}{\sqrt{2\pi}}\sqrt{\uptau}}]$ to the form $b_0\log \uptau +a_0 +a_1 \uptau^{-1/2}+a_2\uptau^{-1}$ for the values $\uptau=1024,1152,1280,1408,1536,1664,1792$ yields $b_0=0.24999920$, and including more subleading corrections improves this even further.

Another heuristic derivation of Eq.~(\ref{eq:iso_asym}) goes as follows. Consider the Fredholm determinant $\det(I-e^{-\varepsilon}V)$, where $\varepsilon>0$. As is explained in \ref{app:alternative}, 
$\det(I-e^{-\varepsilon}V_\alpha)=\det(I- W_\varepsilon)$, where $W_\varepsilon$ now acts on $L^2([0,\sqrt{\uptau}\,])$ with Hankel transform-like kernel
\begin{equation}\label{eq:hankeltransform}
 W_\varepsilon(s,s')=\sqrt{ss'}\int_0^\infty q J_0(sq)J_0(s'q)e^{-q^2/2-\varepsilon}dq.
\end{equation}
For $\varepsilon>0$ the asymptotics follow from Theorem 1.1 in Ref.~\cite{BasorEhrhardt2003}. They are given by
\begin{equation}\label{eq:thm_asymp}
 \det(I-W_{\varepsilon})\underset{\uptau\to\infty}{\sim} \exp\left(b_\varepsilon(0)\sqrt{\uptau}+\frac{1}{2}\int_0^\infty x(b_\varepsilon(x))^2dx\right),
\end{equation}
where $\hat{b}_\varepsilon(q)=\log(1-e^{-q^2/2-\varepsilon})$, and $b_\varepsilon(x)=\int_0^\infty \frac{dq}{\pi}\hat{b}_\varepsilon(q)\cos (qx)$ the inverse cosine transform. For $\varepsilon=0$ the hypotheses of the theorem are not satisfied, due to the singularity at $q=0$. However, it is reasonable to expect the leading term proportional to $\sqrt{\uptau}$, which is still well defined, $\lim_{\varepsilon\to 0}b_\varepsilon(0)=\zeta(3/2)/\sqrt{2\pi}$, to still stay correct. The analogy with standard theory of Toeplitz determinants, where similar pointwise singularities only affect subleading terms \cite{FisherHartwig,gFisherHartwig_proof}, makes this assumption highly plausible. Also, the second term in (\ref{eq:thm_asymp}) diverges logarithmically in the limit $\varepsilon\to 0$, consistent with Eq.~(\ref{eq:iso_asym}).

Let us finally mention that the scaling with $\sqrt{\uptau}$ is expected. Indeed, the problem studied here is closely related to the SSEP, which is described in the continuum limit by a diffusion equation \cite{Spohnbook}, where $\uptau$ plays the role of (real) time for the underlying classical dynamics.  
 \section{Asymptotic results in real time}
 \label{sec:real}
 In this section we switch to real time. The analytic continuation of the partition function (squared) becomes the  return probability after a quench from a domain wall initial state, as was explained in the introduction. Our aim is to determine the long time behavior of the probability and use this to gain some insights into the quench. 
 
 Before getting to the specifics let us mention that the most obvious guess for such long time behaviors would be to bluntly make the substitution $\uptau=it$ in all the results of the previous subsection. There is of course no a priori justification for this, namely asymptotic expansion and analytic continuation have no reason to commute. Also, the result would contradict the overlap argument depicted in section \ref{sec:returnproba}. As we shall see shortly, such a procedure works in certain cases, but fails rather spectacularly in others.
 \subsection{Nowhere continuous behavior with anisotropy}
 \label{sec:fractal}
We now substitute $\uptau=it$ in our exact result (\ref{eq:fred}), and evaluate the resulting determinant numerically using the method explained in \ref{sec:powerseries}. In practice, we were able to access times of the order of $t \approx 500$ to essentially arbitrary numerical precision on a regular laptop. Let us first look at the root of unity case, namely $\Delta=(r+r^{-1})/2$ when $r$ is a root of unity. In that case we parametrize $\gamma$ as
 \begin{equation}\label{eq:rootparametrisation}
\gamma={\rm arccos} \,\Delta=\frac{\pi p}{q},
 \end{equation}
for some coprime integers $p$ and $q$. We observe the formula
\begin{equation}\label{eq:fractal}
 -\log \mathcal{R}(t)=\left(\frac{q^2}{(q-1)^2}-1\right) \frac{(t \sin\gamma)^2}{12}-\kappa(\frac{\pi}{q})\log t +O(1)
\end{equation}
works extremely well\footnote{For denominators up to $q=7$ the formula is accurate up to relative error less than $10^{-4}$ for the times we managed to reach. Finite-size effects increase with $q$, and decrease with time $t$. We also note that it is very difficult to see this effect using ED techniques \cite{real_time}, as the accessible times are not large enough.
}. This result coincides with the analytic continuation of the expansion (\ref{eq:imag_as}), but, surprisingly, only when the integer $p$ in the parametrization (\ref{eq:rootparametrisation}) is set to one. Said differently, the leading long time behavior of the RP, as a function of $t\sin \gamma$, is only sensitive to the denominator in (\ref{eq:rootparametrisation}). Note also the bigger the denominator, the smaller the coefficient of $t^2$ in (\ref{eq:fractal}). Since irrational numbers are limits of rational numbers with increasing numerators and denominators, the coefficient of $t^2$ becomes zero away from root of unity, indicating a change of scaling behavior. We observe the RP decays exponentially in that case, $-\log \mathcal{R}(t)=a(\gamma)t+o(t)$ (the data appears compatible with the simple guess $a(\gamma)=\sin \gamma$). Note also that strictly speaking, a slightly greater scaling such as $t\log t$ or $t\log\log t$ cannot be excluded. 
\begin{figure}[htbp]
\includegraphics[width=8cm]{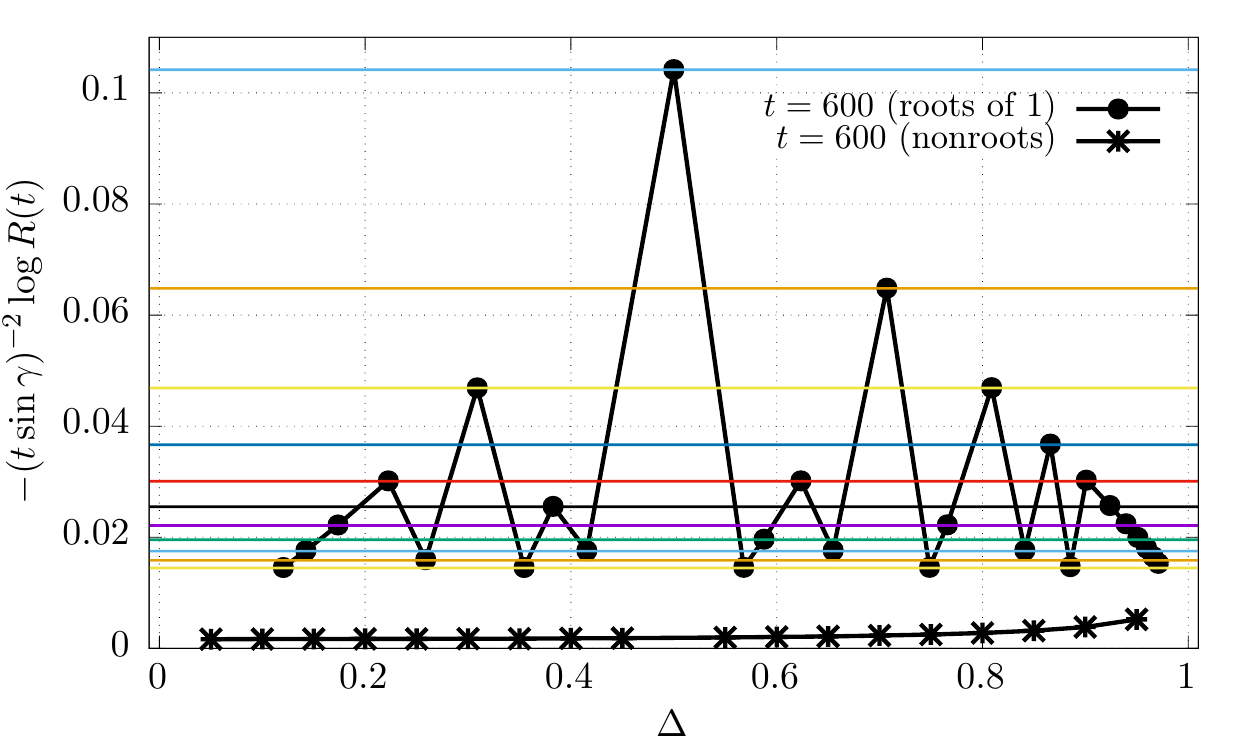}
\includegraphics[width=8cm]{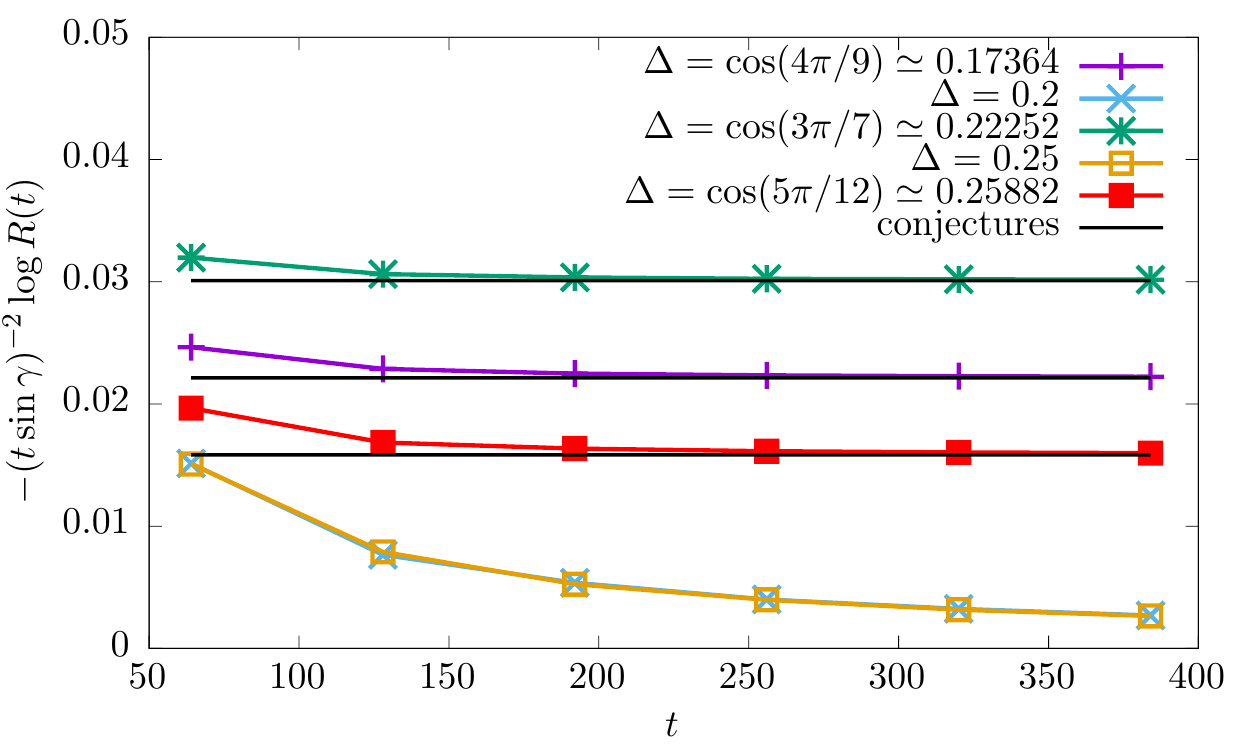}
\caption{Left: roots of unity $\gamma= \pi p/q$ up to denominator $q=13$, and corresponding prefactor of $(t\sin \gamma)^2$ in $-\log \mathcal{R}(t)$ for $t=600$. The straight horizontal lines are the conjectures given by (\ref{eq:fractal}). Note that finite-size effects tend to increase when $\Delta$ approaches $1$. However, we observe that the trend is always good when increasing time. This is illustrated on the right, for a few values of $\Delta$. The results are also improved by including the subleading power law in (\ref{eq:fractal}), as expected (not shown).}
\label{fig:fractal}
\end{figure}
These results are illustrated in figure~\ref{fig:fractal}. The plot allows to emphasize the ``fractal'' nature of the return probability: the coefficient of the gaussian decay is a function of $\Delta$ which is nowhere continuous and evaluates to zero away from root of unity. The effect is quite spectacular when dealing with actual numbers at large times: for example, the return probability at $t=128$ evaluates to approximately $5.90\times10^{-48}$, $2.19\times 10^{-557} $ and $7.04\times10^{-45}$ for $\Delta=0.45$, $\Delta=0.5$ and $\Delta=0.55$ respectively. Spot the odd one out. 

Such a scaling is a clear signature of pathological behavior, stemming from integrability. Our result is reminiscent of known bounds on the high temperature spin Drude weight in the XXZ chain \cite{Zotos,ProsenDrude,IlievskiDrude}, which have been shown to be fractal with $\Delta$ \cite{ProsenDrude,IlievskiDrude}. It is also possible to interpret this result in terms of the string content that is imposed by the domain wall initial state, and which is then used as initial conditions in the generalized hydrodynamic treatment of Refs~\cite{YoungItalians,Doyonhydro}, which is fractal \cite{ColluraDeLucaViti}. It would be interesting to check if this hydrodynamic approach is able to reproduce our result. We also note that for the spin Drude weights, generalized hydrodynamic arguments are also able to reproduce the known fractal bound \cite{IlievskiDeNardisDrude,BerkeleyDrude}. 

It is also unclear which other physical observables might be fractal after the quench, even though the hydrodynamic interpretation suggests that there are. On the other hand, it is not unsurprising that such an effect  be much more visible in the return probability. Indeed, the RP probes an extremely small contribution  to the wave function, which is more likely to show pathological behavior. We leave this question, together with an analytical derivation of (\ref{eq:fractal}) as interesting open problems. We study in the next subsection a simple observable, the speed of the front, which shows continuous behavior with $\Delta$. 
 \subsection{Determination of the front}
 \label{sec:front}
 As already discussed, there is in imaginary time an arctic curve that separates, in the thermodynamic limit, a region where all vertices are frozen from a region where they fluctuate. The real time analog is the front $x_{\rm f}(t)$. For $|x|>x_{\rm f}(t)$ all observables are frozen to their values in the initial domain wall  state, while they fluctuate for $|x|<x_{\rm f}(t)$. This can be simply illustrated at the free fermions point, where the arctic curve is a circle $x^2+y^2=(\uptau/2)^2$. The analytic continuation is obtained \cite{curvedcft} by setting $y=it$ and sending $\uptau\to 0^+$. Hence $x_{\rm f}(t)=t$, and the speed of the front is one, consistent with the exact result (\ref{eq:magn_xx}). 
 
 It is also possible to extract the analytic continuation of the parametric arctic curve (\ref{eq:arcticcurve})  away from free fermions. To do that, write $x/\uptau=f(s)$, $y/\uptau=g(s)$ and suppose we are able to invert the function $g$. Then $x(y)=\uptau f(g^{-1}(y/\uptau))$, and the analytic continuation reads $x_{\rm f}(t)=\lim_{\uptau\to 0^+}\uptau f(g^{-1}(it/\uptau))$.
 Using the asymptotic results (valid for $0<\gamma<\pi/2$)
\begin{eqnarray}
 f\left(\frac{\gamma}{2-\alpha}+i\log z\right)&\,\underset{z\to \infty}{\sim}\,& \frac{\alpha^2(1-\alpha)}{\sin 2\gamma}z^{4-2\alpha}\\
 g\left(\frac{\gamma}{2-\alpha}+i\log z\right)&\,\underset{z\to \infty}{\sim}\,&i\frac{\alpha^2(1-\alpha)}{2\cos \gamma}z^{4-2\alpha},
\end{eqnarray}
we obtain the following equation
\begin{equation}\label{eq:front}
 x_{\rm f}(t)=t\sin \gamma=t\sqrt{1-\Delta^2},
\end{equation}
for the position of the front. The derivation is only valid for $\Delta>0$, but the final result has the correct symmetry when changing $\Delta$ to $-\Delta$ (the quench itself is invariant under this in real time  \cite{SabettaMisguich}, but not in imaginary time). This formula shows excellent agreement with the DMRG results presented in Fig.~\ref{fig:profile}. We note also that the density profile has been computed analytically very recently \cite{ColluraDeLucaViti} using the hydrodynamic approach \cite{YoungItalians}, and reproduces (\ref{eq:front}). The fact that all those results match is strong evidence for the validity of both methods.

Observe also that the speed of the front vanishes in the limit $|\Delta|\to 1$, which means spin transport is sub-ballistic in that case. In the next subsection we argue that it is diffusive.
 \subsection{The special points $|\Delta|=1$}
 \label{sec:diffusion}
\begin{figure}[htbp]
 \centering\includegraphics[width=10cm]{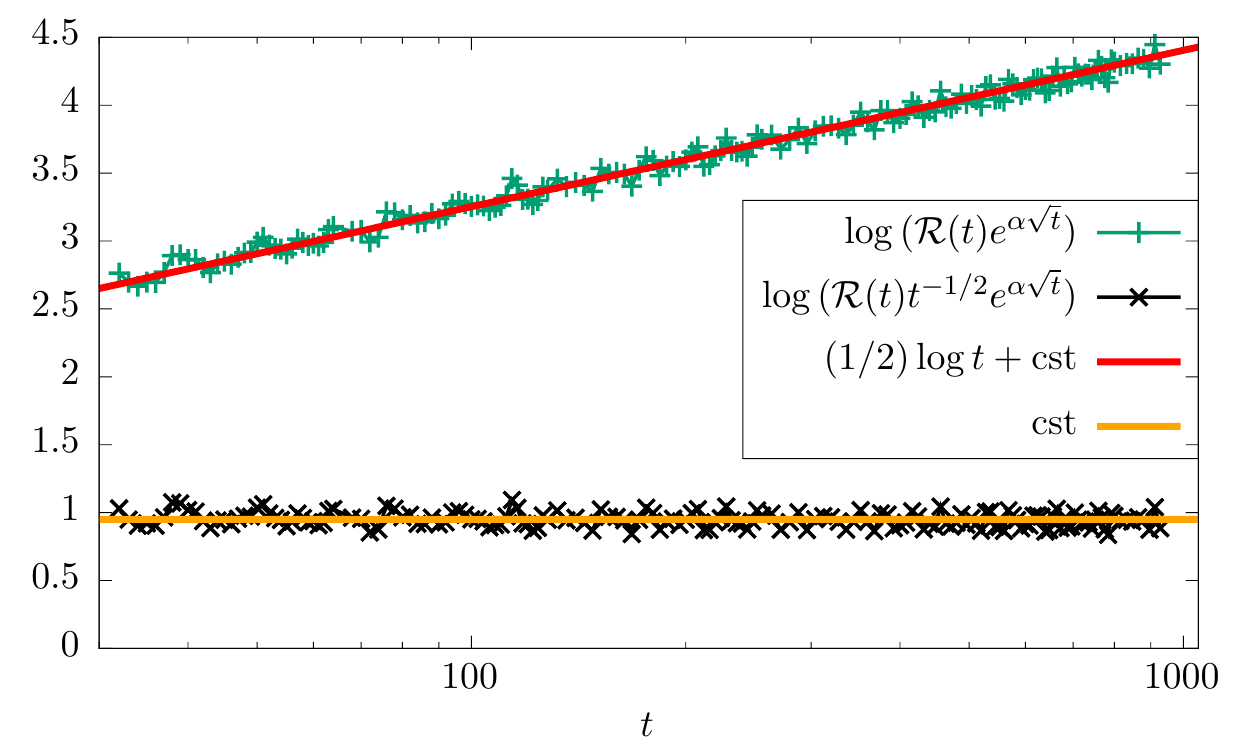}
 \caption{Numerical check of (\ref{eq:xxx_realasympt}) up to time $t=928$. The power law contribution to the RP can also be identified (top curves). Here $\alpha=\zeta(3/2)/\sqrt{\pi}$.}
 \label{fig:root}
\end{figure}

 At the isotropic point (limit $\gamma\to0$, $\gamma\to \pi$ gives the same RP) the return amplitude is given by $\mathcal{A}(t)=\det(I-V)$, with kernel 
\begin{equation}
 V(x,y)=B_0(x,y)e^{i\frac{x+y}{4 t}}.
\end{equation}
The long time behavior may be obtained using the method demonstrated in section~\ref{sec:isotropic}.
We get for the RP
\begin{equation}\label{eq:xxx_realasympt}
 \mathcal{R}(t)=e^{-\zeta(3/2) \sqrt{t/\pi}}t^{1/2}O(1).
\end{equation}
It turns out there is no real subtlety regarding the Wick rotation here, as the result coincides to the leading order with the plain substitution $\uptau=it$ in the imaginary time asymptotic formula (\ref{eq:iso_asym}), and taking modulus square.
A numerical check shown in figure \ref{fig:root} for times up to $t\approx 1000$ shows excellent agreement. Note however that the $O(1)$ term now oscillates, it is not just a numerical constant as in imaginary time. 

It has been recently argued \cite{Prosensuperdiffusion,Prosensuperdiffusion2}, based on DMRG simulations, that spin transport might be super diffusive for this quench, so with a front that would spread as $x_{\rm f}(t)\sim t^{\nu}$, with $\nu \simeq 0.6>1/2$ (an earlier study \cite{Gobert} found $\nu=0.6(1)$). By the logic explained in  section \ref{sec:returnproba}, the overlap with the domain wall initial state would be an overlap between two states with different local properties in a Hilbert space of size exponential in $t^\nu$. Therefore, one would expect $\mathcal{R}(t)\sim e^{-a t^{\nu}}$, which is not consistent with our exact result (\ref{eq:xxx_realasympt}), which does, indeed, suggest diffusive behavior. We note the above argument is of course not a direct proof. In the gapless region, it does predict a return probability that is exponentially small in $t$, which turns out to be incorrect at root of unity. In that case extra perfect cancellations make the return probability much smaller. That is, this argument might heavily \emph{overestimate} the overlap. For superdiffusion to occur at the Heisenberg point, one would need this argument to \emph{underestimate} the overlap. For the sake of completeness, we also study the return probability in the ballistic regime in non integrable (but still $U(1)$) deformations of XXZ in section \ref{sec:nonintegrable}, demonstrating the validity of the Hilbert space size argument in more generic systems. 
 
We also extract numerically the exponent for the spin transport, using a method similar to that of Ref.~\cite{Prosensuperdiffusion}. We study the time-evolution using a tDMRG algorithm, and compute the integrated current $M(t)=\braket{\sum_{x>0} (S_x^3+\frac{1}{2})}_t$, where the average is taken in $\ket{\psi(t)}$. For $\Delta=\pm 1$ we expect a power law $M(t)\sim t^{\alpha}$. 
\begin{figure}[htbp]
 \includegraphics[width=8cm]{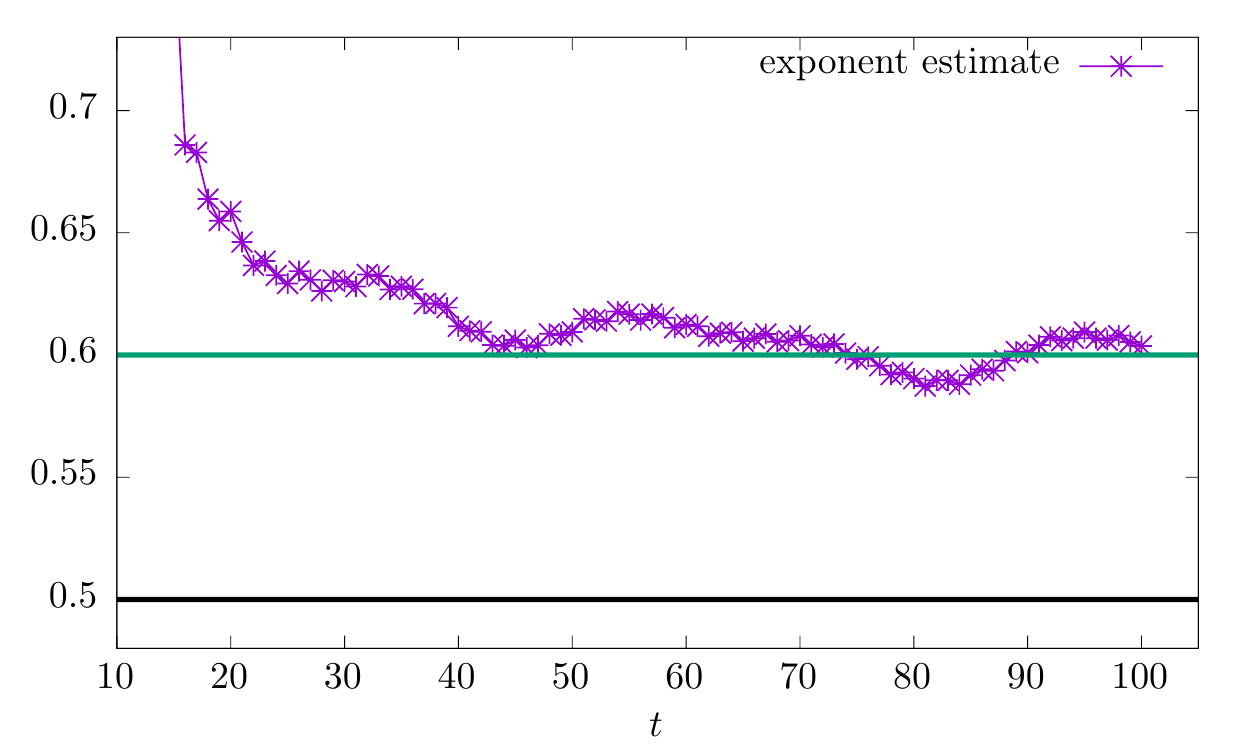}\hfill
 \includegraphics[width=8cm]{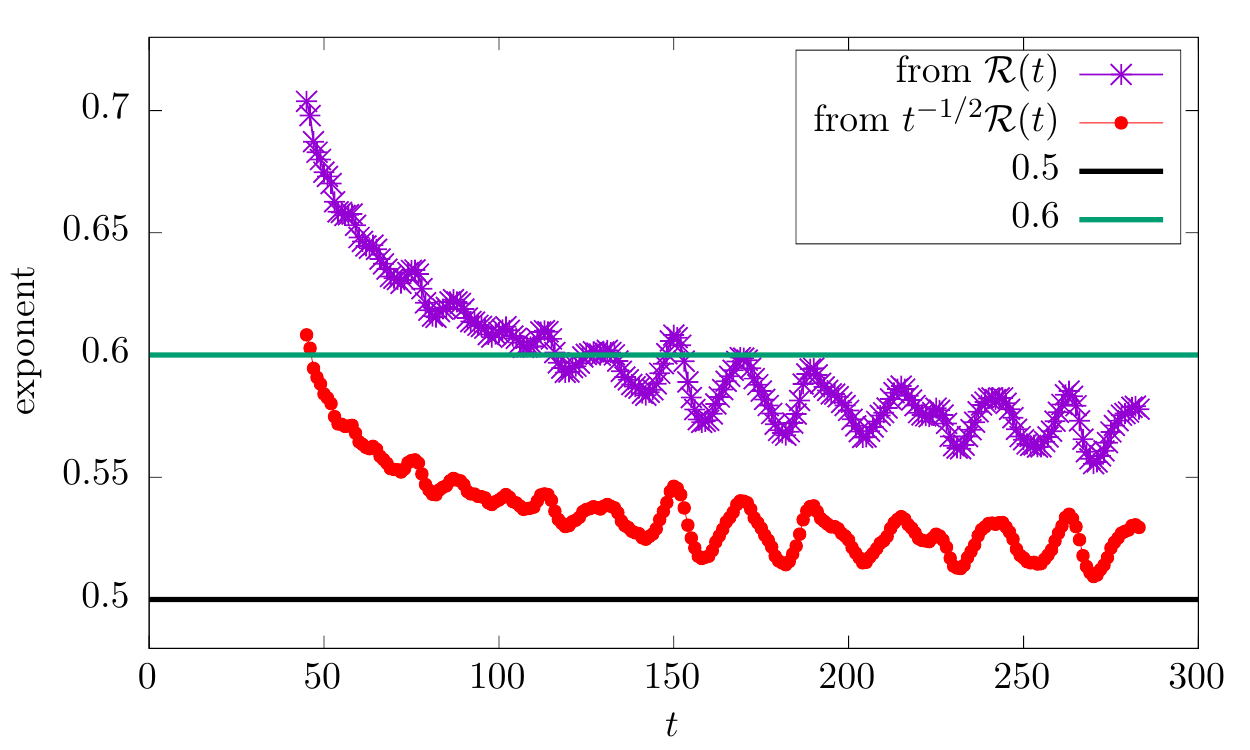}
 \caption{\emph{Left:} Numerical extraction of the transport exponent in the integrated current $M(t)$, up to time $t=100$. We use the method exposed in the text with window $\delta t=15$ (other windows give similar results). Even forgetting about oscillations, it is not so easy to infer the behavior for $t\to\infty$, probably due to slow corrections. \emph{Right:} Numerical extraction of the power law exponent in $\log(\mathcal{R}(t))$ and $\log (t^{-1/2}\mathcal{R}(t))$ assuming (\ref{eq:xxx_realasympt}) is not known, using the same method as on the left with window $\delta t=40$, up to larger times $t\simeq 300$. As can be seen, compensating for the power law correction gives a better estimate of the exponent $\nu=1/2$ in $\mathcal{R}(t)\approx e^{-at^{\nu}}$. }
 \label{fig:transport}
\end{figure}
The simulations were done using the method of Ref.\cite{timeevolvelongrange} which is implemented in the ITensor library \cite{ITensor}. For large times the errors increase, and we chose the truncation in such a way that the return probability be within one percent or less from the exact value. This brings us to about $t\simeq 100$ (which is below what was achieved in Ref.~\cite{Prosensuperdiffusion}). Next, we do a simple fit of $\log M(t)$ to $\alpha\log t+{\rm cst}$ over some window $[t-\Delta t;t]$, and extract $\alpha$, which should give the exponent $\nu$ in the limit $t\to\infty$. The results are shown in figure \ref{fig:transport} on the left. As can be seen, the exponent looks bigger than $0.5$ for the times that are accessible using DMRG. However, the data appears not fully converged, especially since there seems to be a slight trend downwards on top of the inevitable oscillations. 

For comparison, we also applied a similar procedure to the return probability, by fitting $\log(-\log \mathcal{R}(t))$ to the same form $\alpha\log t+{\rm cst}$. In that case oscillations are stronger, and it is necessary to average over a bigger window than for the integrated current. We observe a clear overestimation of the exponent, which almost disappears when getting rid of the power law prefactor in (\ref{eq:xxx_realasympt}). Hence not so small subleading corrections have a significant impact on the extraction of the exponent $\nu$ in this case, and it would not have been so easy to determine the exponent without analytical input. This observation suggests the data for the exponent extracted from the current is also perfectly compatible with $\nu=1/2$. 
\subsection{Non integrable deformations}
 \label{sec:nonintegrable}
 In this section, we look at the same quench from a domain wall initial state in an anisotropic $J_1-J_2$ chain. Our main motivation is to investigate generic behavior away from integrability. The Hamiltonian is given by
 \begin{equation}\label{eq:j1j2}
\fl
 H=\sum_{x}\left(S_x^1 S_{x+1}^1+S_x^2 S_{x+1}^2+\Delta\!\left[S_x^3 S_{x+1}^3-\frac{1}{4}\right]\right)
 +J_2\sum_{x}\left(S_x^1 S_{x+2}^1+S_x^2 S_{x+2}^2+\Delta\!\left[S_x^3 S_{x+2}^3-\frac{1}{4}\right]\right).
\end{equation}
The addition of next nearest neighbor couplings breaks integrability, while still preserving the $U(1)$ symmetry. For $\Delta<1$ and $J_2$ relatively small, one still expects a ballistic spreading of current and correlations, well described by regular hydrodynamic arguments. 
 \begin{figure}[htbp]
 \centering
  \includegraphics[width=10cm]{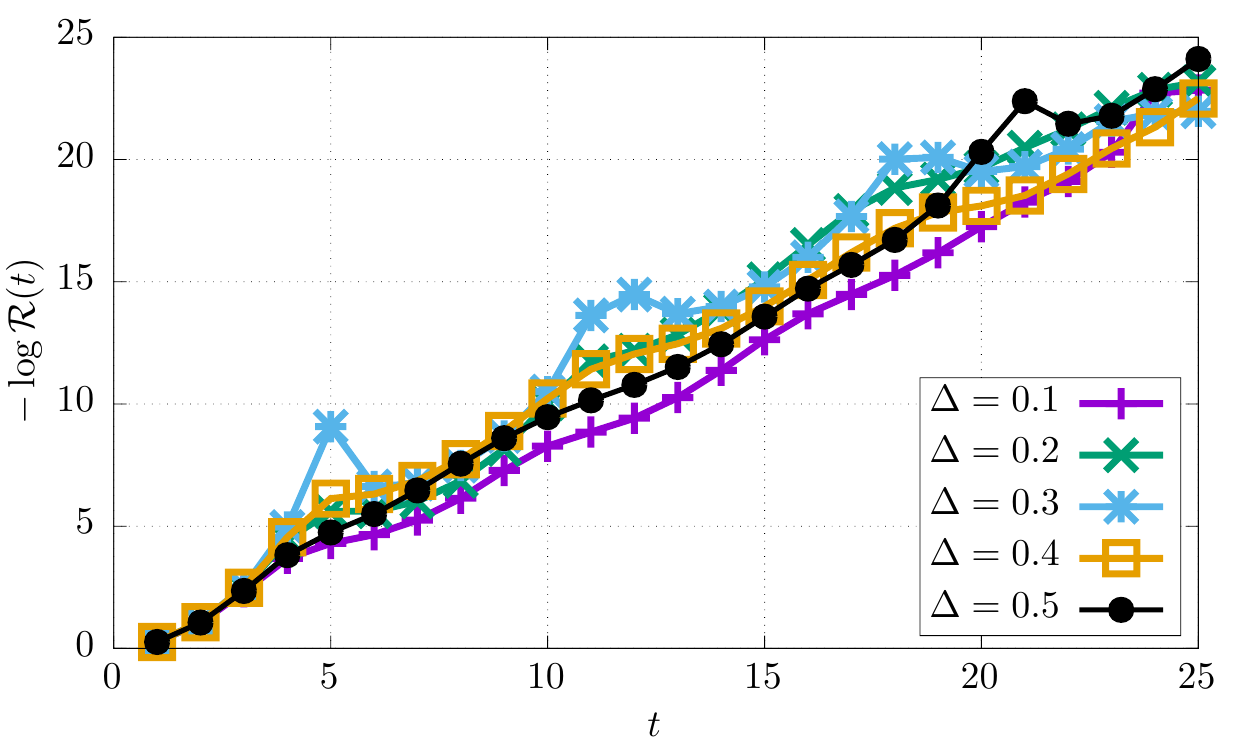}
\caption{Logarithmic return probability as a function of time for several non integrable deformations of the anisotropic XXZ spin chain (\ref{eq:j1j2}) for $J_2=0.16$. We observe a linear scaling with time, apart from some relatively small oscillations.}
\label{fig:nontegrable}
\end{figure}
We focus on the return probability, which is expected to decay exponentially. The integrability breaking terms ensure we avoid the pitfalls encountered at root of unity in the XXZ chain. Here, as can be seen in Fig.~\ref{fig:nontegrable}, the probability decays exponentially fast with $t$ for all values of $\Delta$ we have checked, with the choice $J_2=0.16$. This confirms the validity of the simple Hilbert space counting argument for generic systems with $U(1)$ symmetry. The decay of the RP should be always exponential provided ballistic behavior after the quench. 

We also checked that the ballistic spreading stops for $\Delta\gtrsim 1$. In that case the RP appears to oscillate around some fixed value or decrease very slowly, similar to what happens in the XXZ spin chain.

 \section{Conclusion}
 \label{sec:conclusion}
 In this paper, we investigated the behavior of the return probability after a quench from a domain wall state in the XXZ spin chain. We derived an exact Fredholm determinant formula for it, valid at all times. This was done by using a relation with a certain well-controlled (Trotter, or Hamiltonian) limit of the six vertex model with domain wall boundary conditions, which is also interesting in its own right. The (modulus square of the) Wick rotated partition function is the RP.
 
 In imaginary time the XXZ exhibits the limit shape phenomenon, whereby all degrees of freedom outside of a domain of area $\uptau^2$ are effectively frozen. The boundary of the domain is the arctic curve; we computed it exactly using a result of Colomo and Pronko for the six vertex model \cite{ColomoPronkocurve}. In the limit of large systems, this partition function blows up as $e^{f(\Delta)\uptau^2}$, as expected. We also studied the long time behavior of the RP, and found the following intriguing result. When the anisotropy is at root of unity, the decay of the RP with time is gaussian to the leading order, $e^{-\tilde{f}(\Delta)t^2}$. We conjectured an explicit formula for the rate $\tilde{f}(\Delta)$, which is \emph{nowhere continuous} as a function of $\Delta$, and vanishes away from root of unity (in this case we find an exponential decay with time, with continuous rate). It coincides with $f(\Delta)$ only at principal root of unity $\Delta=\cos \frac{\pi}{q}$ for some integer $q\geq 2$, namely the analytic continuation commutes with the thermodynamic limit only in that case. Such ``fractal'' behavior is due to the integrability of the spin chain: we checked that for non integrable modifications still with $U(1)$ symmetry the generic decay is exponential. 
 
This work raises several questions. The first one is the study of more complicated observables than the return probability, and to determine if other might be also fractal. At this stage, one could argue that such effects should be bigger for the return probability, as more physical observables involve averaging over many components of the wave function. There are probably many others \cite{ColluraDeLucaViti}; for example we observed numerically that the entanglement growth seems slower at root of unity, and so entanglement entropy could be yet another example. We were also able to exhibit one that is not. It is the position of the front, which is given by the simple formula $x_{\rm f}(t)=t\sqrt{1-\Delta^2}$. This result has also a nice interpretation as analytic continuation of known arctic curves in the six vertex model. Exactly at $|\Delta|=1$ we derived the asymptotics of the RP analytically, and found the decay to be consistent with diffusive behavior. Clearly, a better analytical grasp of such transport problems in the Heisenberg model is necessary.

Since we were able to derive an exact determinant formula for it, the return probability is perhaps the simplest observable for which a rigorous long time analysis can be performed. Indeed, similar determinants have been studied using differential equations \cite{JimboMiwaMoriSato,TracyWidom1994}, operator theoretic \cite{BasorEhrhardt2003,EhrhardtBessel}, or Riemann-Hilbert techniques \cite{DeiftZhou,IIKS,BleherFokin,Bothner_RH}. While such methods fall outside the scope of the present paper, a proof of the nowhere continuous behavior, together with the exponential decay away from roots of unity is left as a pressing issue. 
 
 \ack
 This work has benefited greatly from discussions with Filippo Colomo, J\'er\^ome Dubail, Masud Haque, Karol Kozlowski, Olivier Marchal, Gr\'egoire Misguich, Vincent Pasquier and Jacopo Viti. I acknowledge hospitality from the GGI in Florence in 2015, where this project was started.
\paragraph{\bf Note added---} While this paper was being completed, a preprint appeared \cite{ColluraDeLucaViti}, which shows similar fractal behavior for the magnetization and current profile, using the generalized hydrodynamic approach.

 \newpage
 \appendix
 \section[\qquad\qquad\quad Derivation of the determinant formulas]{Derivation of the determinant formulas}
\label{app:fredholmderivation}

Our starting point is the well known formula
\begin{equation}\fl
   Z_n(\epsilon,\gamma)=\frac{\left[\sin \epsilon\right]^{n^2}}{\prod_{k=0}^{n-1}k!^2}\det_{0\leq i,j\leq n-1}
   \left(
M_{ij}(\epsilon)
   \right)\qquad,\qquad M_{ij}(\epsilon)=\int_{-\infty}^{\infty} du \,u^{i+j} e^{-\epsilon u} \frac{1-e^{-\gamma u}}{1-e^{-\pi u}}.
 \end{equation}
 quoted in the main text. 
 Our aim here is to compute the following limit
 \begin{equation}
 \mathcal{Z}(\uptau)=\braket{e^{\uptau H}}=\lim_{n\to \infty} Z_n(\epsilon_n,\gamma)
 \end{equation}
where $\epsilon_n$ satisfies $\frac{\sin \epsilon_n}{\sin(\gamma+\epsilon_n)}=\frac{\uptau}{2n}$. The limit is non trivial because $\epsilon$ goes to zero as $n$ goes to infinity. To do that we will need two results on orthogonal polynomials \cite{Szegobook}, see below.
\subsection[\qquad\qquad\quad Hankel determinants and orthogonal polynomials]{\bf Two results on Hankel determinants and orthogonal polynomials}
\label{sec:ortho}
Let us consider some scalar product $\langle f,g \rangle=\int dx f(x)g(x)w(x)$ for some weight function $w(x)$, and a set of \emph{monic} (i.e. with leading coefficient one, $p_k(x)=x^k+\ldots$) polynomials $\{p_k(x)\}_{k\geq 0}$ orthogonal for this scalar product: $\braket{p_k,p_l}=h_k \delta_{kl}$. Next, consider the $n$ by $n$ Hankel matrix $A=(A_{ij})=\braket{x^{i+j}}$. Then it holds
\begin{equation}\label{eq:det}
   \det A\,=\,\det_{0\leq i,j\leq n-1}(\braket{x^{i+j}})\,=\,\det_{0\leq i,j\leq n-1}\left(\braket{p_{i}p_{j}}\right)
   \,=\,\prod_{k=0}^{n-1}h_{k},
\end{equation}
and \cite{Borodin_biortho}
\begin{equation}\label{eq:inverse}
 (A^{-1})_{ij}=\left.\frac{\partial^{i+j}K_n(x,y)}{i!j!\partial x^i\partial y^j}
 \right|_{
x=y=0
 }
 \quad,\quad 
\end{equation}
with kernel
\begin{equation}
 K_n(x,y)=\sum_{k=0}^{n-1}\frac{p_k(x)p_k(y)}{h_k}=\frac{1}{h_{n-1}}\frac{p_n(x)p_{n-1}(y)-p_{n-1}(x)p_n(y)}{x-y}.
\end{equation}
The last equality on the rhs is called the Christoffel-Darboux formula. In case the polynomials are known explicitly, the above results (\ref{eq:det},\ref{eq:inverse}) provide explicit formulas for the determinant and the inverse of the corresponding Hankel matrix.
\subsection[\qquad\qquad\quad A warming-up exercise]{\bf A warming-up exercise} 
We look at $\lim_{\epsilon\to 0}Z_n(\epsilon,\gamma)$ for fixed $n$. This limit should obviously give one (see  figure \ref{fig:hamlimit}), but let us derive it by a direct calculation. The most singular part in the matrix is given by 
\begin{equation}
E_{ij}=\int_{0}^{\infty}du u^{i+j}e^{-\epsilon u}=\frac{(i+j)!}{\epsilon^{i+j+1}} 
\end{equation}
Now we make use of the results of (\ref{sec:ortho}). The measure (with $\epsilon=1$) is the weight function associated to the well-known Laguerre polynomials, so (\ref{eq:det}) may be applied. To be more precise, the monic orthogonal polynomials associated to the weight $w(x)=e^{-\epsilon x}$ on $\mathbb{R}^+$, are given by  $p_k(x)=(-1)^k\frac{k!}{\epsilon^k}L_k(\epsilon x)=x^k+\ldots$, where $L_k(x)$ is the k-th normalized Laguerre polynomial. The normalization coefficient is then simply $h_k=\int_0^\infty dx \,p_k(x)^2e^{-\epsilon x}=\frac{k!^2}{\epsilon^{2k}}\int \frac{dy}{\epsilon}L_k(y)^2 e^{-y}=\frac{k!^2}{\epsilon^{2k+1}}$. Therefore
\begin{eqnarray}
 \lim_{\epsilon \to 0} Z_n(\epsilon,\gamma)&=&\lim_{\epsilon\to 0} \left\{\frac{[\sin\epsilon]^{n^2}}{\prod_{k=0}^{n-1}k!^2}
\prod_{k=0}^{n-1}\frac{k!^2}{\epsilon^{2k+1}} 
 \right\}\\
 &=&1
\end{eqnarray}
as should be.
\subsection[\qquad\qquad\quad Back to our problem]{\bf Back to our problem}
 Using the result of the previous exercise the imaginary time partition function is given by
 \begin{eqnarray}
  \mathcal{Z}(\uptau)&=&\lim_{n \to \infty} \left\{\left(\frac{\sin \epsilon_n}{\epsilon_n}\right)^{n^2} \frac{\det_{0\leq i,j\leq n-1}\left(M_{ij}(\epsilon_n)\right)}{\det_{0\leq i,j\leq n-1}\left(E_{ij}(\epsilon_n)\right)}
  \right\},
 \end{eqnarray}
where it is understood that $\frac{\sin \epsilon_n}{\sin(\gamma+\epsilon_n)}=\frac{\uptau}{2n}$. The first factor is easy,
\begin{equation}
 \lim_{n \to \infty} \left(\frac{\sin \epsilon_n}{\epsilon_n}\right)^{n^2}=e^{-(\uptau\sin \gamma)^2/24},
\end{equation}
so we are left with $\mathcal{Z}(\uptau)=e^{-(\uptau\sin \gamma)^2/24}\tilde{\cal Z}(\uptau)$, with
\begin{equation}
 \tilde{\cal Z}(\uptau)=\lim_{n \to \infty} \left\{\frac{\det_{0\leq i,j\leq n-1}\left(M_{ij}(\epsilon_n)\right)}{\det_{0\leq i,j\leq n-1}\left(E_{ij}(\epsilon_n)\right)}
  \right\}
\end{equation}
Now we invert the matrix $E$, using formula (\ref{eq:inverse}), and get
\begin{eqnarray}\nonumber
 (E^{-1}M)_{ij}&=&\delta_{ij}-\sum_{k=0}^{n-1}
 \left.\frac{\partial^{i+k}K_{n}(x,y)}{i!k!\partial x^i \partial y^k}\right|_{x,y=0}
\int_{-\infty}^{\infty} dz \,z^{k+j} e^{-z\epsilon_n}\left(\Theta(z)-\frac{1-e^{-\gamma z}}{1-e^{-\pi z}}\right)\\
&=&\delta_{ij}-\left.\frac{\partial^i}{i!\partial x^i}\int_{-\infty}^\infty dz K_{n}(x,z)z^j e^{-z\epsilon_n}\left(\Theta(z)-\frac{1-e^{-\gamma z}}{1-e^{-\pi z}}\right)\right|_{x=0}
\end{eqnarray}
The kernel can be written down in terms of Laguerre polynomials. We have
\begin{eqnarray}
 K_{n}(x,y)&=&\frac{1}{h_{n-1}}\frac{p_n(x)p_{n-1}(y)-p_{n-1}(x)p_n(y)}{x-y}\\
 &=&\frac{\epsilon_n^{2n-1}}{(n-1)!^2}\frac{-n!(n-1)!}{\epsilon_n^{n}\epsilon_n^{n-1}}\frac{L_n(\epsilon_n x)L_{n-1}(\epsilon_n y)-L_{n-1}(\epsilon_n x)L_n(\epsilon_n y)}{x-y}\\
 &=&-n \frac{L_n(\epsilon_n x)L_{n-1}(\epsilon_n y)-L_{n-1}(\epsilon_n x)L_n(\epsilon_n y)}{x-y}
\end{eqnarray}
Now the limit $n\to \infty$ can be taken, and the Laguerre kernel reduces to a Bessel kernel ($T=\uptau \sin \gamma$): 
\begin{equation}
 K(x,y)=\frac{\sqrt{2yT}J_0(\sqrt{2xT})J_0'(\sqrt{2yT})-\sqrt{2xT}J_0(\sqrt{2yT}) J_0'(\sqrt{2xT})}{2(x-y)}
\end{equation}
Hence the infinite determinant representation
\begin{equation}
 \tilde{\mathcal{Z}}(\uptau)=\det_{i,j}\left(\delta_{ij}-
 \left.\frac{\partial^i}{i!\partial x^i}\int_{-\infty}^\infty dz K(x,z)z^j\left(\Theta(z)-\frac{1-e^{-\gamma z}}{1-e^{-\pi z}}\right)\right|_{x=0}
 \right)
\end{equation}
for which we use the lighter notation
\begin{equation}\label{eq:infinitedet}
 \tilde{\mathcal{Z}}(\uptau)=\det_{i,j}(\delta_{ij}-V_{ij})=\det_{i,j}\left(\delta_{ij}-
 \left.\frac{\partial^i}{i!\partial x^i}\int_{-\infty}^\infty dz V(x,z)z^j\right|_{x=0}
 \right),
\end{equation}
with $V(x,z)=K(x,z)\left(\Theta(z)-\frac{1-e^{-\gamma z}}{1-e^{-\pi z}}\right)$. Eq.~(\ref{eq:infinitedet}) can also be rewritten as a Fredholm determinant. Indeed, since $\log \det(.)= {\rm Tr}\,\log (.)$, we have
\begin{eqnarray}\fl
 \log \tilde{\mathcal{Z}}&=&\log \det_{i,j}(\delta_{ij}-V_{ij})
 \\\nonumber\fl
 &=&-\sum_{k=1}^{\infty} \frac{1}{k} \sum_{i_1,\ldots ,i_k=0}^\infty V_{i_1 i_2}V_{i_2 i_3}\ldots
 V_{i_k,i_1}\\\nonumber\fl
 &=&-\sum_{k=1}^{\infty} \frac{1}{k} \sum_{i_1,\ldots ,i_k=0}^\infty 
 \int_{-\infty}^{\infty} dz_1\ldots dz_k
 \left.\frac{\partial^{i_1}}{\partial x_1^{i_1}}\ldots
  \frac{\partial^{i_k}}{\partial x_k^{i_k}}
  V(x_1,z_2)V(x_2,z_3)\ldots V(x_k,z_1)\frac{z_1^{i_1}}{i_1!}\ldots \frac{z_{k}^{i_k}}{i_k!}\right|_{x_1=0,\ldots, x_k=0}\\\nonumber\fl
  &=&-\sum_{k=1}^{\infty} \frac{1}{k} \int_{-\infty}^{\infty} dz_1\ldots dz_k V(z_1,z_2)V(z_2,z_3)\ldots V(z_k,z_1)\\\nonumber\fl
  &=& -\sum_{k=1}^\infty \frac{{\rm tr}\, V^k}{k}\\\fl\label{eq:somefred}
  &=&\log \det (I-V)
\end{eqnarray}
Putting everything back together one gets the Fredholm with kernel
\begin{equation}\fl
 V(x,y)=\frac{\sqrt{2yT}J_0(\sqrt{2xT})J_0'(\sqrt{2yT})-\sqrt{2xT}J_0(\sqrt{2yT}) J_0'(\sqrt{2xT})}{2(x-y)}\left[\Theta(y)-\frac{1-e^{-\gamma y}}{1-e^{-\pi y}}\right],
\end{equation}
where recall $T=\uptau\sin\gamma$. By simple change of variables, the determinant formula (\ref{eq:somefred}) also holds with kernel
\begin{equation}\fl
 V(x,y)=\frac{\sqrt{y}J_0(\sqrt{x})J_0'(\sqrt{y})-\sqrt{x}J_0(\sqrt{y}) J_0'(\sqrt{x})}{2(x-y)}\omega(y)\qquad,\qquad \omega(y)=\Theta(y)-\frac{1-e^{-\frac{\gamma y}{2\uptau \sin \gamma}}}{1-e^{-\frac{\pi y}{2\uptau \sin \gamma}}},
\end{equation}
which is the result claimed in section \ref{sec:hamilimit}. Note the two interesting limits: $\gamma\to 0^+$ ($\Delta\to 1^-$), which gives $\omega(y)=e^{-y/2\uptau}\Theta(y)$, and $\gamma\to \pi^-$ ($\Delta\to -1^+$) which gives $\omega(y)=-e^{y/2\uptau}\Theta(-y)$.
\subsection[\qquad\qquad\quad Another determinant representation,  and the case $\Delta>1$]{\bf Another determinant representation,  and the case $\Delta>1$}\label{app:another}
The strategy used to derive a Fredholm determinant for $\mathcal{Z}(\uptau)$ relied on a formula of the type $Z_n=\det A/\det B=\det(B^{-1}A)$, finding an explicit expression for the matrix elements of $B^{-1}$, and finally taking the Hamiltonian limit. Since $B$ is a Hankel matrix whose entries are moment of Laguerre polynomials, the inverse follows from (\ref{eq:inverse}) by plugging the explicit expressions of the Laguerre polynomials. This gives a determinant in terms of a Laguerre kernel, which becomes a Bessel kernel after taking the Hamiltonian limit. 

As pointed out in Ref.~\cite{ColomoPronkodet}, it is also possible to use Meixner-Pollaczek instead of Laguerre polynomials, which provides an alternative representation for the partition function $Z_n$. Taking the Hamiltonian limit, we find
\begin{equation}
 \mathcal{Z}(\uptau)=e^{-i\frac{\uptau}{2} \sin \gamma}\det(I-V),
\end{equation}
\begin{equation}
 V(x,y)= e^{i(\uptau \sin \gamma-\gamma)}\frac{f_\uptau(x)g_\uptau(y)-f_\uptau(y)g_\uptau(x)}{x-y} \frac{e^{2\gamma y}}{1+e^{2\pi y}},
\end{equation}
with
\begin{equation}
 f_\uptau(x)\;=\; _{1\!}F_1(ix+1/2,1,-i\uptau \sin \gamma)\qquad,\qquad g_\uptau(x)=-\uptau\partial_\uptau f_\uptau(x),
\end{equation}
and
\begin{equation}
 _{1\!}F_1(a,b,z)=\sum_{k=0}^{\infty} \frac{(a)_k}{(b)_k}\frac{z^k}{k!}\qquad,\qquad (c)_k=c(c+1)\ldots (c+k-1)=\frac{\Gamma(c+k)}{\Gamma(c)}
\end{equation}
is Kummer's confluent hypergeometric function. 

This result turns out to be useful to treat the gapped phase $\Delta>1$. In that case $\gamma=i \eta$, $\cosh \eta=\Delta$ becomes pure imaginary, and the multiple integrals defining the Fredholm determinant may be computed as discrete sums using the Residue theorem. We obtain
\begin{equation}\label{eq:deltagreateronedet}
 \mathcal{Z}(\uptau)=e^{\frac{\uptau}{2} \sinh \eta}\det_{0\leq j,l\leq \infty}\left(\delta_{jl}-V_{jl}\right),
\end{equation}
\begin{equation}
 V_{jl}=\frac{u_\uptau(j)v_\uptau(l)-v_\uptau(j)u_\uptau(l)}{j-l}e^{-\uptau \sinh \eta} e^{-\eta (j+l)},
\end{equation}
with
\begin{equation}
 u_\uptau(j)\;=\;_{1\!}F_1(-j,1,\uptau \sinh \eta)\qquad,\qquad v(j)=\uptau \partial_\uptau u_\uptau(j).
\end{equation}
The large $\uptau$ behavior follows from the observation that $u_\uptau(j)$ and $v_\uptau(j)$ are polynomials of degree $j$ in $\uptau$, which means off-diagonal elements of $V_{jl}$ decay exponentially fast with $\uptau$. Using $V_{jj}=e^{-\uptau\sinh \eta -2j \eta}\left.\left(\frac{d u_\uptau(x)}{dx}v_\uptau(x)-\frac{d v_\uptau(x)}{dx}u_\uptau(x)\right)\right|_{x=j}$, we get $V_{00}=1- e^{-\uptau \sinh \eta}$ and $V_{jj}\sim e^{-2\eta j}$ for $j\geq 1$. Computing the determinant as the product of its diagonal elements yields
\begin{equation}\label{eq:deltagreaterone}
 \mathcal{Z}(\uptau)\sim e^{-\frac{\uptau}{2} \sinh \eta}\prod_{k=0}^{\infty}\left(1-e^{-2k\eta}\right).
\end{equation}
The real time case was already studied in \cite{MosselCaux}, using different techniques. It was shown analytically that $\overline{\mathcal{R}}(t)\geq \prod_{k=1}^\infty\left(1-e^{-2k\eta}\right)^2$ \footnote{Note that Ref.\cite{MosselCaux} studied a large periodic chain, in which case the return probability is squared compared to our setup. This is simply due to the presence of two domain walls $\ldots\uparrow\uparrow\uparrow\downarrow\downarrow\downarrow\ldots$ and $\ldots\downarrow\downarrow\downarrow\uparrow\uparrow\uparrow\ldots$ in a periodic system.} for a time averaged RP, together with compelling numerical evidence that the bound is tight. This coincides with the analytic continuation of (\ref{eq:deltagreaterone}), which can be seen as an alternative heuristic derivation. One could of course try to obtain a rigorous proof by making the analytic continuation in (\ref{eq:deltagreateronedet}) first, and then performing an asymptotic expansion. However the structure of the determinant becomes much more complicated in that case, and this task turns out to be not straightforward. We leave this as an interesting problem. 
\subsection[\qquad\qquad\quad Alternative representations at $\Delta=1$]{\bf Alternative representations at $\Delta=1$}
\label{app:alternative}
We provide here several equivalent representations for $\mathcal{Z}(\uptau)$ when $\Delta=1$. All these follow from repeated use of the identity $-\log \det(I-V)=\sum_{k=1}^\infty \frac{1}{k}{\rm Tr}\, V^k$ with
\begin{equation}
 {\rm Tr}\, V^k =\int dx_1\ldots dx_k V(x_1,x_2)V(x_2,x_3)\ldots V(x_k,x_1),
\end{equation}
the change of variable $\det(I-V)=\det(I-\tilde{V})$, with kernel $\tilde{V}(x,y)=\sqrt{\frac{df}{dx}\frac{df}{dy}}V(f(x),f(y))$, and changing the order of integration. We start from the formula $\mathcal{Z}(\uptau)=\det(I-\tilde{V})$, with
\begin{equation}
 \tilde{V}(x,y)=\frac{1}{4}e^{-(x+y)/4}\int_0^\uptau ds J_0(\sqrt{sx})J_0(\sqrt{s y})
\end{equation}
[The identity $\tilde{V}(x,y)=\uptau V(\uptau x,\uptau y)$ where $V(x,y)$ is given by (\ref{eq:fred_delta1}) may be established by taking the derivative with respect to $\uptau$ on both sides of the equality.] We have
\begin{equation}\fl
 {\rm Tr}\, \tilde{V}^k=\int_0^\infty \frac{dx_1}{4}\ldots \frac{dx_k}{4}\int_0^\uptau ds_1\ldots ds_k e^{-\frac{x_1}{2}-\ldots -\frac{x_k}{2}}
 J_0(\sqrt{s_1 x_1})J_0(\sqrt{s_1 x_2})\ldots J_0(\sqrt{s_k x_k})J_0(\sqrt{s_k x_1})
\end{equation}
Changing the order of integration we get ${\rm Tr}\, \tilde{V}^k={\rm Tr}\, W^k$, with kernel
\begin{equation}\label{eq:somebesselint}
 W(s,s')=\int_0^\infty dx e^{-x/2}J_0(\sqrt{sx})J_0(\sqrt{s'x})
\end{equation}
acting on $L^2([0,\uptau])$. Making the change of variable $x\mapsto x^2$ in the previous integral, as well as $s\mapsto s^2$, $s'\mapsto s'^2$ in the kernel yields the kernel (\ref{eq:hankeltransform}) acting on $L^2([0,\sqrt{\uptau}])$. Using $J_0(u)=\int_{0}^{2\pi} \frac{d\theta}{2\pi}e^{i u\cos \theta}$, (\ref{eq:somebesselint}) may also be rewritten as
\begin{equation}
 W(s,s')=\frac{1}{2}e^{-(s+s')/2}I_0(\sqrt{ss'}),
\end{equation}
where $I_0(v)=\int_0^{2\pi} \frac{d\theta}{2\pi}e^{v\cos \theta}$ is a modified Bessel function. It also admits the contour integral representation
\begin{equation}
 W(s,s')=\oint_{|z|=r} \frac{dz}{4i\pi z}e^{\frac{s}{2}(z-1)+\frac{s'}{2}(1/z-1)},
\end{equation}
where the contour is any circle of radius $r>0$. Hence
\begin{equation}
 {\rm Tr}\, W^k=\int_0^{\uptau/2}ds_1\ldots ds_k \oint \frac{dz_1}{2i\pi z_1}\ldots \frac{dz_k}{2i\pi z_k} e^{s_1(1/z_1+z_2-2)+\ldots+s_k(1/z_k+z_1-2)},
\end{equation}
which coincides exactly with the antepenultimate equation in appendix B of \cite{DerridaGerschenfeld2009} ($\uptau=2t$ in that reference, and $\omega=-1$, which corresponds to vanishing current).
 \section[\qquad\qquad\quad Numerical evaluation of Fredholm determinants]{Numerical evaluation of Fredholm determinants}
\label{app:fredholmnumerical}
We provide here some information regarding the numerical evaluation of the determinants formulas for the return probability. We deal here with determinants of the form
\begin{equation}\label{eq:fredholmdetnotation}
 \det(I-V),
\end{equation}
where $V$ is an operator acting on $L^2([a,b])$ for some choice of $a$ and $b$. Either may be infinite if need be. The Fredholm determinant (\ref{eq:fredholmdetnotation}) is then defined as
\begin{equation}\label{eq:fredholm_def}
 \det(I-V)=\sum_{n=0}^{\infty} \frac{1}{n!} \int_{a}^b dx_1\ldots \int_{a}^b d x_n \det_{1\leq i,j\leq n} \left(V(x_i,x_j)\right),
\end{equation}
or, alternatively, through its logarithm
\begin{equation}
 \log \det (I-V)=-\sum_{n=1}^{\infty} \frac{1}{n}{\rm Tr}\, V^n,
\end{equation}
with
\begin{equation}
 {\rm Tr}\, V^n =\int dx_1 \ldots dx_n V(x_1,x_2)V(x_2,x_3)\ldots V(x_{n-1},x_n)V(x_n,x_1).
\end{equation}
We discuss two ways of evaluating such determinants. One is to rely on power series expansion of the kernel (\ref{sec:powerseries}), the other is quadrature \cite{Bornemann} (\ref{sec:quadrature}).
\subsection[\qquad\qquad\quad Power series method]{\bf Power series method}
\label{sec:powerseries}
It is possible to trade the multiple integrals defining ${\rm Tr}\, V^n$ for discrete sums, by power series expansion of the kernel around some point (any fast converging series also works). Injecting $V(z_\alpha,z_\beta)=\sum_{i=0}^{\infty}\left. \frac{\partial^{i}}{i ! \partial x}V(x,z_\beta)z_\alpha^{i}
\right|_{x=0}$ into the definition of the determinant, we obtain the ``infinite'' determinant representation
\begin{eqnarray}\label{eq:powerseriesdetmethod}
 \det (I-V)&=&\det_{0\leq i,j\leq \infty}\left(\delta_{ij}-\frac{\partial^i}{i! \partial x^i}\left.\int_a^b  V(x,z) z^j \,dz \right|_{x=0} \right)\\
 \label{eq:fullyexplicit}
 &=&\det_{0\leq i,j\leq \infty}\left(\delta_{ij}-\sum_{k=0}^{\infty} \frac{\beta_{k+j+1}(\uptau)}{(i+k+1)j!^2k!^2} \right),
\end{eqnarray}
with 
\begin{equation}\label{eq:betaq}\fl
 \beta_q=\left(\frac{\uptau \sin \gamma}{2\pi}\right)^q (q-1)!\left[((-1)^{q}+1)\zeta(q)+(-1)^{q+1}\zeta(q,\gamma/\pi)-\zeta(q,1-\gamma/\pi)\right] \quad ,\quad q\geq 2,
\end{equation}
where $\zeta(q,a)=\sum_{n\geq 0}(a+n)^{-q}$ is the Hurwitz zeta function, $\zeta(q)=\zeta(q,1)$, and $\beta_1(\uptau)=\frac{\uptau \cos \gamma}{2}$.  
This method is clearly inspired by the results of \ref{app:fredholmderivation}, where identity (\ref{eq:powerseriesdetmethod}) was an intermediate step in the derivation of the Fredholm determinant. The formula is useful, due to the following two facts: (i) the matrix elements can be computed efficiently, see (\ref{eq:fullyexplicit},\ref{eq:betaq}), and (ii) accurate numerical evaluation may be performed by only keeping the first $n$ rows and columns in the determinant. For example in real time $\uptau=it$, we found that it is only necessary to have $n/t$ constant to achieve exponentially fast convergence for large $t$ (e.g. a ratio $n/t\approx 3$ appears to be always sufficient). Hence reasonably large times may be accessed. The only disadvantage is that the resulting matrix is extremely ill-conditioned, which means it is necessary to compute the matrix elements with very high accuracy, e. g. by using a computer algebra system. In practice, the number of necessary digits also scales linearly with time, resulting in a $t\times t^3=t^4$ complexity for the determinant evaluation (and $t\times t^2=t^3$ memory consumption). 

We were able to reach values of the order of $t\approx 500$ or slightly larger on a regular laptop, the main bottleneck being memory. We note also that the need for high accuracy is not so surprising, since the final result is expected to be extremely small, especially at roots of unity. For example, the return amplitude for $t=256$ at $\Delta=1/2$ is $\sqrt{\mathcal{R}(t)}\simeq 6.14246\times 10^{-1113}$. This is not true anymore at $\Delta=1$, where the decay is much slower. In that case evaluation of the Fredholm determinant by quadrature methods discussed below outperforms the infinite determinant method.  
\subsection[\qquad\qquad\quad Quadrature method]{\bf Quadrature method}
\label{sec:quadrature}
For $\Delta=1$ the imaginary and real time kernels may be rewritten as
\begin{equation}\label{eq:numerical_kernels}
 V_\uptau(x,y)= B_\uptau(x,y) e^{-\frac{x+y}{2}}\qquad,\qquad W_t(x,y)=B_t(x,y) e^{i\frac{x+y}{2}},
\end{equation}
with 
\begin{equation}
 B_T(x,y)=\frac{\sqrt{2Ty}J_0(\sqrt{2T x})J_0'(\sqrt{2Ty})-\sqrt{2Tx}J_0(\sqrt{2T y})J_0'(\sqrt{2Tx})}{2(x-y)},
\end{equation}
where both kernels act on $L^2(\mathbb{R}_+)$. We will use both examples to illustrate the use of quadrature methods to compute the determinant.
\paragraph{Reminder on gaussian quadrature methods for integrals ---}
Say we wish to compute numerically a one-dimensional integral of the form
\begin{equation}
 I=\int_a^b \,dx f(x) w(x),
\end{equation}
where $w(x)$ is a known weight, and $f(x)$ is a smooth function. Such integrals are usually computed by discretization,
\begin{equation}\label{eq:int_discretization}
 I\simeq \sum_{j=1}^n w_j f(x_j).
\end{equation}
There is a huge literature on the determination of best discretization abscissas and weights, according to certain criteria. One particularly efficient method dates back to Gauss. The aim is to find, for each choice of $n$ (number of function evaluation), a ``best'' possible choice of abscissas $(x_j)_{1\leq j\leq n}$ and weights $(w_j)_{1\leq j\leq n}$, such that the approximation (\ref{eq:int_discretization}) be exact when $f(x)$ is a polynomial of degree $2n-1$ or less. Since smooth function can be well approximated by polynomials, such methods are extremely efficient. 

The abscissa and weights may be determined using orthogonal polynomials techniques. 
One starts by constructing a set of \emph{monic} polynomials $\{p_k(x)\}_{k\geq 0}$, orthogonal for the scalar product
\begin{equation}
 \int_a^b p_k(x)p_l(x) w(x)dx=h_k \delta_{kl}.
\end{equation}
A fundamental property of such orthogonal polynomials is that they obey a three-term recurrence relation
\begin{equation}
 p_{k+1}(x)=(x-\alpha_k)p_k(x)-\beta_k p_{k-1}(x).
\end{equation}
One uses these coefficients to build the Jacobi matrix
\begin{equation}
 J_n=\left(\begin{array}{cccccc}
        \alpha_0&\sqrt{\beta_1}    &0&0&\ldots&0\\
        \sqrt{\beta_1}&\alpha_1&\sqrt{\beta_2}&0&\ldots&0\\
          0&\sqrt{\beta_2}&\alpha_2&\sqrt{\beta_3}&\ldots&0\\
          \vdots&\ldots&\ddots&\ddots&\ddots&\vdots\\
        0&\ldots&0&\sqrt{\beta_{n-2}} &\alpha_{n-2}&\sqrt{\beta_{n-1}}\\
        0&0&0&0&\sqrt{\beta_{n-1}}&\alpha_{n-1}
           \end{array}
\right)
\end{equation}
Then, one can show that the correct abscissas $x_j$ are the eigenvalues of the Jacobi matrix, and the weights are given by $w_j=\frac{h_{n-1}}{p_{n-1}(x_j)p'_n(x_j)}$. They may also be determined from the eigenvectors of the Jacobi matrix: denoting by $(v_{jl})_{1\leq l\leq n}$ the eigenvector corresponding to the eigenvalue $x_j$, we have $w_j=h_0 (v_{j1})^2$, $h_0=\int_a^b w(x)dx$.

Such quadrature methods are much easier to implement if the recurrence coefficients are known exactly, which is a rare occurrence. Useful examples include the Gauss-Legendre quadrature ($w(x)=1$ on $[-1,1]$), the Gauss-Hermite quadrature ($w(x)=e^{-x^2}$ on $\mathbb{R}$), the Gauss-Laguerre quadrature ($w(x)=e^{-x}$ on $\mathbb{R}_+$), and the Gauss-Jacobi quadrature, which generalizes the Gauss-Legendre to weights of the form 
 $w(x)=(1-x)^a(1+x)^b$, $a,b>-1$ on $[-1,1]$.
 \paragraph{Oscillating integrals that decay slowly---}
Let us now consider a Fourier type integral, of the form
\begin{equation}
 \hat{F}(w)=\int_0^\infty dx f(x) e^{i\omega x},
\end{equation}
for some slowly decaying function $f(x)$. For this type of integrals the method exposed in the previous paragraph does not work, due to the oscillations combined with the slow decay of the integrand. A clever ``double-exponential'' quadrature method to treat such integrals has been put forward in Refs.~\cite{OouraMori,Ooura}. First, one makes the following change of variables
\begin{equation}
 \varphi(u)=\frac{u}{1-\exp\left[-2u-\alpha (1-e^{-u})-\beta(e^u-1)\right]}
\end{equation}
which maps $\mathbb{R}_+$ onto $\mathbb{R}$. The function $\varphi(u)$ behaves as $\varphi(u)\sim u$ for large positive $u$ and vanishes double exponentially fast for large negative $u$. After the change of variable, a trapezoidal rule with mesh size $h$ is applied to the integral, resulting in the approximation
\begin{equation}\label{eq:dequad}
 \hat{F}(\omega)\simeq \frac{\pi}{\omega}\sum_{p=-\infty}^{\infty} f\left(\frac{\pi \varphi(ph)}{\omega h}\right)\varphi'(ph)
 \left[e^{\frac{i\pi}{h}\varphi(ph)}+(-1)^{p+1}\right].
\end{equation}
The choice of parameters $\beta=1/4$, $\alpha=\beta\left[1+\log\left(1+\frac{\pi}{\omega h}\right)/(4\omega h)\right]^{-1/2}$ is advocated in Ref.~\cite{OouraMori}. For typical smooth slow-decaying functions $f(x)$ the quadrature formula (\ref{eq:dequad}) converges exponentially fast (as $e^{-A/h}$ for some constant $A$) to the exact value. Another crucial feature is that numerical evaluation may be performed by simple truncation of the sum. Indeed, for large negative $p$, $\varphi'(ph)$ decays double exponentially fast to zero, while for large positive $p$, $\varphi(ph)\simeq ph$ up to double exponentially small corrections, which means the last factor in square bracket in (\ref{eq:dequad}) is double exponentially small. Hence we truncate the sum as $\sum_{p=-\infty}^{\infty}\simeq \sum_{p=-P_-}^{P_+}$ and choose both $P_{\pm}$ to be proportional to $1/h$. 
\paragraph{Application: Fredholm determinant computations at the isotropic point---} As pointed out by Bornemann \cite{Bornemann}, gaussian-type quadrature methods may be used to compute Fredholm determinant of smooth kernels. The kernel $V_\uptau$ of (\ref{eq:numerical_kernels}) is a good example. Due to the exponential decay of the kernel for large arguments, it is very natural to use Gauss-Laguerre quadrature. Denoting by $x_i$ and $w_i$ the abscissa and weights corresponding to the $n$-point quadrature, the Fredholm determinant is simply approximated by performing the quadrature on the multiple integral representation (\ref{eq:fredholm_def}). We obtain
\begin{equation}
 \det(I-V_\uptau)\,\simeq\, \det_{1\leq i,j\leq n}\left(\delta_{ij}-w_j B_\uptau(x_i,x_j)\right),
\end{equation}
or, in a symmetric form,
\begin{equation}
 \det(I-V_\uptau)\,\simeq\, \det_{1\leq i,j\leq n}\left(\delta_{ij}-\sqrt{w_i w_j}\, B_\uptau(x_i,x_j)\right)
 \end{equation}
 For smooth kernels this converges exponentially fast in $n$ to the exact value \cite{Bornemann}, which means extremely accurate computations may be performed with a reasonable number of quadrature points. For large $\uptau$, we observed that the number of quadrature points scales linearly with $\uptau$; despite this limitation, we managed to reach values $\uptau \approx 2000$ or even higher. 
 
The case of slowly decaying oscillating kernels such as $W_t$ is trickier, as there is no general result regarding the convergence of quadrature methods on (\ref{eq:fredholm_def}). It is however very natural to try and implement the double exponential quadrature method \cite{Ooura} exposed in the previous paragraph, with $\omega=1$. Using the following set of abscissas and weights
\begin{equation}
 x_p=\frac{\pi \varphi(ph)}{h}\qquad,\qquad w_p=\pi\varphi'(ph) \left[e^{\frac{i\pi}{h}\varphi(ph)}+(-1)^{p+1}\right],
\end{equation}
we found the approximation
\begin{equation}
 \det(I-V)\,\simeq\, \det_{-P_-\leq p,q\leq P_+}\left(\delta_{pq}-w_q B_t(x_p,x_q)\right)
\end{equation}
to be extremely accurate. For example, the choice of parameters $P_-=7M/4$, $P_+=3M/2$, $M=\left \lfloor{7t/4}\right \rfloor $, $h=\pi/M$, $\alpha=0.25$, $\beta=0.125$ yields the correct value to machine precision in the range $t\in [64;500]$ (for $t$ smaller it is necessary to make $M$ slightly larger). In that case  we were able to compare to the (slower) infinite determinant method. It also allows to go further, e. g. in Fig.~\ref{fig:root} some data up to $t=928$ is shown. 

Let us stress that we only applied this method on the Fredholm determinant we were interested in, as well as a few other simple examples, were it appears to work extremely well also. It is therefore unclear to us what is the accuracy of this procedure in full generality. Here, the choice of parameters was justified a posteriori by comparing to the results of the infinite determinant method, an exact result for the integral $\int_0^\infty V_{t}(x,x)dx=te^{it}[i J_0(t)-J_1(t)]/2$, as well as the asymptotic expansion (\ref{eq:xxx_realasympt}). 
 \section*{References}
\bibliography{Return.bib}{}
 \bibliographystyle{h-physrev}
\end{document}